\documentclass[11pt]{article}
\usepackage[square,authoryear]{natbib}
\usepackage{marsden_article}

\def\Re{{I\!\!R}}

\newcommand{\D}{{\cal H}}
\newcommand{\I}{{\cal I}}
\newcommand{\hO}{\hat{\Omega}}
\newcommand{\hT}{\hat{T}}
\newcommand{\ba}{\bar{a}}
\newcommand{\bB}{\bar{B}}
\newcommand{\hA}{\hat{\alpha}}

\newcommand{\g}{\cal G}

\begin{document}

{\huge {\bf

\title{Nonholonomic systems  via moving frames:\\ Cartan equivalence and Chaplygin Hamiltonization}
\author{Kurt Ehlers\\
Department of Mathematics \\
Truckee Meadows Community College\\
7000 Dandini Blvd Reno NV  89512-3999  USA \\
{\footnotesize kehlers@tmcc.edu} \and
Jair Koiller\\
Funda\c c\~ao Getulio Vargas\\ Praia de
Botafogo 190,Rio de Janeiro, 22253-900 Brazil \\
{\footnotesize jkoiller@fgv.br}
\and
Richard Montgomery \\
Mathematics Department \\ 
University of California at Santa Cruz \\
Santa Cruz CA 95064 USA \\
{\footnotesize rmont@math.ucsc.edu}
\and Pedro M. Rios\\
Department of Mathematics\\ University of California  at Berkeley\\ Berkeley CA 94720 USA.\\
{\footnotesize prios@math.berkeley.edu}\\
}


\date{{\footnotesize submitted, October, 2003; revised version, April, 2004}\footnote{The authors  thank the  
 Brazilian funding agencies CNPq and FAPERJ: a CNPq  research fellowship (JK), 
a CNPq post-doctoral fellowship at Berkeley (PMR),  a   FAPERJ visiting fellowship
to Rio de Janeiro (KE). (JK) thanks the E. Schr\"odinger Institute, Vienna, 
for financial support during Alanfest and the Poisson Geometry Program, August 2003.}.\\[12pt]  }

\bigskip
\bigskip

\maketitle

}}

{\bf {\large

\centerline{{\it Dedicated to Alan Weinstein on his 60th Birthday}}

}}


\newpage

{\Large {\bf
\centerline{Abstract}
}}


 A {\it nonholonomic  system}, for short ``NH'', consists of a configuration space $Q^n$, a Lagrangian $L(q,\dot{q},t)$,  a nonintegrable constraint  distribution ${\cal H} \subset TQ$, with dynamics   governed by Lagrange-d'Alembert's principle. We present here two   studies,  both using {\it adapted moving frames}.   In the first   we explore  the {\it affine connection viewpoint}. 
For natural Lagrangians $ L = T - V$, where we take $V=0$ for simplicity, 
NH-trajectories   are   geodesics of a (non metric) connection  $\nabla_{NH}$ 
which mimics  Levi-Civita's.    
   {\it Local geometric invariants} are obtained by Cartan's  method of equivalence.  
As an example, we analyze Engel's (2-4) distribution.   This is the first such study for a distribution that is not strongly nonholonomic.
In the second part we study $G$-{\it Chaplygin  systems}; for those,  
 the constraints are given by a connection $\phi: TQ \rightarrow Lie(G)$ on a principal bundle $G \hookrightarrow Q \rightarrow S = Q/G$ and the Lagrangian $L$ is   G-equivariant. 
These systems compress to an almost Hamiltonian system 
$(T^*S, H^{\phi},\Omega_{NH})$,  
$\,\,\Omega_{NH} = \Omega_{can} + (J.K)$, with $d(J.K) \neq 0 $ in general;     the momentum map $\,J:T^*Q \rightarrow Lie(G)$ and
  the curvature form $K: TQ \rightarrow Lie(G)^*$ are matched via the Legendre transform.  Under a  $s \in S$ dependent
time reparametrization, a number of compressed systems become Hamiltonian, i.e, $\Omega_{NH}$
 is sometimes conformally symplectic.  A  necessary condition is the existence of an
invariant volume for the original system. Its density produces a candidate for conformal factor. Assuming an invariant volume, we describe the obstruction to Hamiltonization. An example of
Hamiltonizable system is the ``rubber'' Chaplygin's sphere, which extends Veselova's system
in $T^*SO(3)$. This is a ball with unequal 
inertia coefficients rolling without slipping on the plane, with vertical rotations forbidden.  Finally, 
 we discuss  reduction of internal symmetries.
Chaplygin's ``marble'', where vertical rotations
are allowed,   is not Hamiltonizable 
at the compressed $T^*SO(3)$ level. We conjecture that  it is also {\it not} Hamiltonizable when {\it reduced} to $T^*S^2$.   

\tableofcontents

\newpage

{\small
\begin{quotation} ``Nonholonomic mechanical systems (such as systems with rolling contraints)
provide a very interesting class of systems where the reduction procedure has to be modified. In fact this provides a class
of systems that give rise to an almost Poisson structure, i.e, a bracket which does not
necessarily satisfy the Jacobi identity  
'' (\cite{MW}).
\end{quotation}
}

\section{Introduction and outline}

Cartan's moving frames method is a standard tool in Riemannian geometry\footnote{ \cite{Cartan0}; there is a recent
 English translation from the Russian translation  (\cite{Cartan1}).  One of the most important  applications was the construction of characteristic classes by   Alan's adviser, S.S. Chern. Our taste for moving frames in Mechanics is a small tribute to his influence.}.
  In Analytical Mechanics, the method  goes back to \cite{Poincare},
 perhaps earlier, to Euler's  rigid body equations, perhaps  much earlier, to the
caveperson who invented the wheel.   Let $q \in \Re^n$ be local coordinates on  a configuration space $Q^n$, 
and consider a local  frame, defined by  an $n \times n$ invertible matrix $B(q)$,
\begin{equation}
X_j = \frac{ \partial}{\partial \pi_j} = \sum_{i=1}^n\, b_{ij} \frac{ \partial}{\partial q_i}\,\,\,\,,\,\,\,\, \sum \,\dot{\pi}_j\, X_j = \sum \, \dot{q}_i \frac{ \partial}{\partial q_i}\,\,,\,\, \dot{\pi} = A(q) \dot{q}\,\,,\,\,\,A = B^{-1} \,\,.
\end{equation}
In  Mechanical Engineering, \cite{Hamel}, \cite{Papastavridis},  moving frames 
disguise under the keyword {\it quasi-coordinates}, nonexisting entities $\pi$ such that
$$ \frac{ \partial f}{\partial \pi_j} = \sum_i\, \frac{ \partial f}{\partial q_i}\, \frac{ \partial q_i}{\partial \pi_j} = \sum_i\, \frac{ \partial f}{\partial q_i}\, b_{ij} = X_j(f)
$$
Let $\{\epsilon_i\}_{i=1,...,n}$ be the dual coframe to  $\{ X_j \}$,
 $\epsilon_i = {\rm ``}\,d\pi_i {\rm ''} = \sum_j a_{ij} \,dq_j\,\, .$

\subsection{Moving frames: Lagrangian and Hamiltonian  mechanics } 
The Euler-Lagrange 1-form  rewrites  
as\footnote{Atributed to Hamel, but certainly known by Poincar\'e. Quasi-coordinates can be found in \cite{Whittaker} and were first used in Mechanics by Gibbs, see \cite{Pars}.}:  
\begin{equation}  \label{Hamel}
\sum_{r=1}^n \,\left( \frac{d}{dt}\frac{\partial L}{\partial \dot{q}_r} - \frac{\partial L}{\partial q_r} - F_r \right)\, dq_r = \sum_{k=1}^n\,\, \left( \frac{d}{dt}\frac{\partial L^*}{\partial \dot{\pi}_k} - \frac{\partial L^*}{\partial \pi_k} + \,\sum_{i=1}^n\, \frac{\partial L^*}{\partial \dot{\pi}_i} \sum_{j=1}^n   \gamma^i_{kj} \dot{\pi}_j - R_k \right) \, \epsilon_k  = 0
\end{equation}
where 
$ L^*(q,\dot{\pi},t) = L(q,B(q)\dot{\pi},t)$ is the Lagrangian written in 
``quasi-coordinates'' and $R_k = \sum_s\, F_s\,b_{sk}$ are the covariant components
 of the total force (external, $F_{ext}$, and  constraint force $\lambda$). 
The so called {\it Hamel's transpositional symbols}
$\gamma^i_{kj} = \ \gamma^i_{jk} = \sum_{s,\ell = 1}^n \, b_{sk}\,b_{\ell j} \left( \partial a_{is}/\partial q_{\ell} - \partial a_{i \ell}/\partial q_s \right)$
 are precisely the moving frame structure coefficients  (\cite{KoillerARMA}).

 If  the velocities are restricted to a subbundle ${\cal H} \subset
TQ$, a {\it constraint force} $\lambda$   appears. {\it D'Alembert-Lagrange principle}\footnote{According to \cite{Sommerfeld}, this gives the most natural foundation for Mechanics.} 
  implies that $\lambda$  belongs to the anihilator ${\cal H}^o  \subset T^*Q$ of ${\cal H}$,
 hence exerting zero work on admissible motions $\dot{q} \in {\cal H}$:
\begin{equation}
[L] := \frac{d}{dt}\, \frac{\partial L}{\partial \dot{q}} - \frac{\partial L}{\partial q} - F_{ext} = \lambda \in {\cal H}^o \,\,, \,\, \dot{q} \in {\cal H} \,\,.
\end{equation}

Using moving frames,   constraints can be eliminated directly.
 If ${\cal H}^o $ is spanned by
the last r forms $\epsilon_J,\,  s + 1 \leq J \leq n \,\,\, (s = n - r)$, then   equations
of motion result from setting  the first $s$ Euler-Lagrange differentials equal to zero:
\begin{equation}  \label{Hamel1}
  \frac{d}{dt}\frac{\partial L^*}{\partial \dot{\pi}_k} - \frac{\partial L^*}{\partial \pi_k} + \,\sum_{i=1}^n\, \frac{\partial L^*}{\partial \dot{\pi}_i} \sum_{j=1}^n   \gamma^i_{kj} \dot{\pi}_j - F^{ext}_k  = 0\,\,\,\,\,\,\,\,\,\,\,\,\,\, (1 \leq k \leq s) \,\,.
  \end{equation}
  
Strikingly, {\it the Hamiltonian counterparts
 of }(\ref{Hamel}) and (\ref{Hamel1})
{\it are  simpler, although less known}\footnote{A ``moving frames operational system'' for Hamiltonian mechanics
 in $T^*Q$ was given in  \cite{Koillerromp}.}. The philosophy    is
to fight against Darboux's dictatorship.  In terms of the  local coframe  
$ \{ \, \epsilon_i \}_{1 \leq i=1 \leq n}$,   any element $p_q \in T^*Q$ can be written as
$ p_q = \sum \,m_i\,\epsilon_i(q)$.  The natural 1-form $\alpha $  on $T^*Q$ keeps the  familiar confusing expression
$ \alpha : = pdq = m \epsilon \,.\,\,$
 Consequently, the canonical symplectic form $\Omega : = d\alpha$
 writes as 
\begin{equation} \label{mepsilon}
\Omega : = dp \wedge dq = dm \wedge \epsilon + m \,d\epsilon \,\,.
\end{equation}
The second term $m \,d\epsilon$, which deviates from Darboux's format, is {\it not} a nuisance,
 it  carries   most valuable information. For instance,  Kostant-Arnold-Kirillov-Souriau's 
bracket in $T^*G$, $G$ a Lie group, can be immediately visualized: take a
 (left or right) invariant coframe and apply H. Cartan's ``magic formula'' on $d\epsilon$. So moving frames are ideally suited when a Lie symmetry group $G$ is present\footnote{As we learned 
from Alan  at the banquet, the etymology for {\it symplectic}
is ``capable to join'', themes and people. The latter is one of the most 
important aspects of the symplectic ``creed''. {\it Provocation}: taking moving frames, 
{\it adapted} to some other
mathematical structure for $Q$, would the non-Darboux term provide  a
 {\it local} symplectic invariant?}.

\paragraph{Example: mechanics in $SO(3)$.}  To fix notation, we now review the standard example.
  The Lie algebra basis $X_i \in sO(3) = T_I SO(3),\,i=1,2,3$ (infinitesimal rotations
around the $x,y,z$-axis at the identity),   can  either be right or
left transported, producing   moving frames on $SO(3)$ denoted $\{X_i^{r}\}$
and $\{X_i^{\ell}\}$ respectively. Let $\{\rho_i\}_{1 \leq i\leq 3}$ and
$\{\lambda_i\}_{1 \leq i \leq 3}$, denote  
their dual coframes (right and left invariant forms in $SO(3)$).  To represent angular momenta,  we use Arnold's notations (\cite{Arnold1}): capital letters mean
objects in body frame,  smallcase objects in the space frame. Thus for instance,  $\ell = R L$, where $L$ is the angular momentum in body frame
and $\ell$ is the angular momentum in space; likewise $\omega = R \Omega$ relate
the angular velocities. The canonical 1-form in $T^*SO(3)$ is given by
$$ \alpha = \ell_1 \rho_1 + \ell_2 \rho_2 + \ell_3 \rho_3 = L_1 \lambda_1 + L_2 \lambda_2 + L_3 \lambda_3$$ so 
$$ \Omega_{can} = \sum d\ell_i \rho_i   + \ell_1 d\rho_1 + \ell_2 d\rho_2 + \ell_3 d\rho_3 =  \sum dL_i \lambda_i +
L_1 d\lambda_1 + L_2 d\lambda_2 + L_3 d\lambda_3,
$$
where by Cartan's structure equations, $d\lambda_1 = - \lambda_2 \wedge \lambda_3, \cdots$
and $d\rho_1 = \rho_2 \wedge \rho_3, \cdots$ (cyclic). A  left invariant metric  is given by an inertia operator $L = A \Omega$. Euler's rigid body equations follow immediately. 

\paragraph{Poisson action of $S^1$ on $SO(3)$.} Consider the {\it left} $S^1$ action on $SO(3)$ given by   $\exp(i \phi) \cdot R  :=  S(\phi) \, R$ 
where $S(\phi)$ is the rotation matrix about the $z$-axis:
$$
 S(\phi) := \left( \begin{array}{lll} \cos(\phi) & -\sin(\phi) & 0 \\
                                           \sin(\phi) & \cos(\phi) & 0 \\
                                           0 & 0 & 1 \end{array} \right) \,\,,\,\,\,
 S(-\phi) S'(\phi) = \left( \begin{array}{lll} 0 & -1 & 0 \\
                                        1 & 0 & 0 \\
                                           0 & 0 & 0 \end{array} \right) \,\, = X_3 \,\, .   
$$
Two matrices are in the same
equivalence class iff their third rows, which we denote by $\gamma$,  called the {\it Poisson vector},     are the same: 
$ R_1 \sim R_2  \Longleftrightarrow  R_1^{-1} \hat{k} =   R_2^{-1} \hat{k} = \gamma \in S^2\,\,.$
So we have a principal bundle $\pi:SO(3) \rightarrow S^2$, 
$  \gamma = \pi (R) = R^{-1} \hat{k} = R^{\dagger}\, \hat{k}  $.
  The derivative of $\pi$ is 
\begin{equation} \label{1/2}
\dot{\gamma} = \pi_*(\dot{R}) = - (R^{-1}\dot{R}R^{-1}) k = - (R^{-1}\dot{R}) (R^{-1}) k = - [\Omega] \gamma = - \vec{\Omega} \times \gamma =  \gamma \times \vec{\Omega}
\end{equation}
where we used the custumary identification\footnote{We will drop the $[\bullet]$ and $\vec{\bullet}$ in the sequel, and mix  all the notations, hoping no confusion will arise. Equation (\ref{1/2}) is one half of every system of ODEs for $S^1$-equivariant
mechanics in $SO(3)$.  Of course, we also obtain $\dot{\gamma} = - \Omega \times \gamma$ by differentiating 
 $R \gamma = k$ (we could use the notation
$\gamma = K$, but we won't).}
 $[\Omega] \in sO(3) \leftrightarrow \vec{\Omega} \in \Re^3$, \cite{Arnold1}. The lifted action
to $T^*SO(3)$ has momentum map  $J = \ell_3$.

 \paragraph{Connection on  $S^1 \hookrightarrow SO(3) \rightarrow S^2$.}   Take the usual bi-invariant metric $<<\,,\,>>$ on $SO(3)$ so that both $\{X_i^{\ell}\}$ and $\{X_i^{r}\}$
are orthonormal moving frames. The tangent vectors to the fibers are  $(d/d\phi) S(\phi)\cdot R = X_3^{right}$.  Consider the {\it mechanical connection} associated to $<<\,,\,>>$, namely,
horizontal and vertical spaces are orthogonal. The horizontal spaces are generated by $X_1^{right}$ and $X_2^{right}$. The connection form is $\phi = \rho_3$.  The horizontal lift of $\dot{\gamma}$ to $R$ is the tangent vector $\dot{R}$ such that 
 \begin{equation} \label{hor}
\Omega_{hor} = R^{-1} \dot{R} = [ \dot{\gamma} \times \gamma ]
\end{equation}
Note that $\Omega_{hor}$ {\it is the -90 degrees rotation of } $\dot{\gamma}$ {\it inside} $T_{\gamma} S^2$. 
The curvature of this connection $\kappa = d\rho_3$ is the area form of the sphere.

\paragraph{Reduction of $S^1$ symmetry.} It is   convenient for reduction 
 to use  $(a,\ell_3), \, a \in \Re^3, \, a \perp \gamma \,$,
\begin{equation} \label{Lb}
      L := a \times \gamma + \ell_3 \gamma
\end{equation}
where  $a$ is a vector perpendicular to $\gamma$.  The vector $a$ has an intrinsic meaning:
Consider  a moving frame $e_1, e_2$ in $S^2$, with dual coframe
$\theta_1, \theta_2$. Then $\,\, v_{\gamma} = v_1 e_1 + v_2 e_2 $ parametrizes $TS^2$, and $\,\,p_{\gamma} = a \cdot d\gamma = p_1 \theta_1 + p_2 \theta_2$ parametrizes $T^*S^2$,\,
$a = p_1\,e_1 + p_2\, e_2$.
Here  $a \cdot d\gamma\,,\,\, \sum \gamma_i d\gamma_i = 0$   denotes both an element of $T^*S^2$ and the canonical 1-form.  Our parametrization for $SO(3)$ is $\,\, R(\phi,\gamma) = S(\phi)\cdot R(\gamma),\, R(\gamma) = {\rm rows} (e_1, e_2, \gamma)$.  Then $L = p_2\,e_1 - p_1\,e_2 + \ell_3\,\gamma$ corresponds to $\ell = (p_2,- p_1, \ell_3)$ {\it  along the section} $\phi = 0$. The right invariant forms are compactly represented as 
\begin{equation} \label{compactly}
\rho_3 = d\phi - (de_1,e_2)\,\,,\,\, \rho_1 + i \rho_2 = -i \exp(i\phi) (\theta_1 + i \theta_2)\,\,.
\end{equation}
Lifting $v \in TS^2$ to an horizontal
vector in $TSO(3)$ is simple: 
\begin{equation} \label{horSO3}
\Omega_{hor}  = [ (v_1\, e_1 + v_2\, e_2 ) \times \gamma ] = [v_2\,e_1 - v_1\,e_2] \,\,\,{\rm or} \,\,\, hor(v) = v_2\, X_1^r - v_1\,X_2^r\,\,,
\end{equation}
Hence any vector $\dot{R} \in
TSO(3)$ can be written as $\dot{R} = \omega_1\,X_1^{\ell} +  \omega_2\,X_2^{\ell} + \omega_3\,X_3^{\ell}$
with $\omega_1 = v_2, \omega_2 = - v_1$. 
Any covector $p_R \in
T^*SO(3)$ can be written as $p_R = p_1\, \pi^*(\theta_1) + p_2\, \pi^*(\theta_2) + \ell_3\,\rho_3$. 

The reduced symplectic manifold $J^{-1}(\ell_3)/S^1 \equiv T^*S^2$ can be {\it explicitly} constructed,  taking the section $\phi = 0$. Let   $i: T^*S^2 \rightarrow T^*SO(3)$,
\begin{equation} \label{sectioni}
 i(\gamma,p_1,p_2) = (R(\gamma), \ell)\,,\,\, \ell = (p_2,-p_1,\ell_3) \,\,.
\end{equation}
 Then from (\ref{compactly}) we get
$i^*\,\rho_2 = -  \theta_1 \,,\,\, i^*\,\rho_1 =  \theta_2$, and $i^*d = di^*$ yields
$$i^*\,d\rho_1 = d\theta_2\,,\,\,i^*\,d\rho_2 = - d\theta_1\,,\, i^*\, d\rho_3 = i^*\rho_1 \wedge i^*\rho_2 = - \theta_2 \, \theta_1 = \theta_1 \, \theta_2 .$$
We get immediately
\begin{equation} \label{redSO3}
 \Omega^{red}_{T^*S^2}  =  i^*(\Omega_{T^*SO(3)}) = d(p_1\,\theta_1 + p_2\,\theta_2) +      \ell_3 \,{\rm area} = \Omega^{can}_{T^*S^2} + \ell_3 \,{\rm area}_{S_2} \,\,.
\end{equation}
All references to the moving frame disappear,  but the expression $\Omega^{can}_{T^*S^2} = d(p_1\,\theta_1 + p_2\,\theta_2)$, suggests that whenever a natural mechanical system in $T^*SO(3)$ reduces to $T^*S^2 \equiv TS^2$, there is
a prefered choice for the moving frame $\{e_1,e_2\}_{\gamma}$:  namely, that one which diagonalizes the Legendre transform  $ T_{\gamma}\,S^2 \rightarrow T^*_{\gamma}\,S^2 \equiv T_{\gamma}\,S^2$
of the reduced (Routh) Lagrangian. 

 \subsection{Nonholonomic systems} 
 A NH system $(Q,L,{\cal H})$  consists of a configuration 
space $Q^n$,   a Lagrangian $L: TQ \times \Re \rightarrow \Re$ , and a totally nonholonomic constraint distribution   ${\cal H} \subset TQ $. 
The dynamics are governed by Lagrange-d'Alembert's principle\footnote{ ``Vakonomic'' mechanics uses  the same ingredients, but the dynamics are governed by the variational principle with constraints, and produce different equations, see e.g. 
\cite{CortesManolo}. The equations coincide if and only if the distribution is integrable.
In spite of many similarities, 
there are striking 
differences between NH and holonomic systems.  For instance,  NH systems do not 
have (in general) a smooth invariant measure. Necessary and sufficient conditions 
 for the existence of the invariant measure were first given (explicitly in coordinates) by  \cite{Blackall}.}. Usually $L$ is {\it natural}, $L = T - V$ where $T$is the kinetic energy associated to a Riemannian metric $\langle\,,\,\rangle$, and $V = V(q)$
 is a potential.
By {\it totally nonholonomic} we mean that the filtration
$ {\cal H} \subset {\cal H}_1 \subset {\cal H}_2 \subset ... $
ends in $TQ$. Each sub-bundle ${\cal H}_{i+1}$ is obtained from the previous one by 
adding to ${\cal H}_i$  combinations of all possible Lie brackets of   vectorfields
 in ${\cal H}_i$. To avoid interesting complications we assume  that all have constant rank.
 Equivalently, let ${\cal H}^o \subset T^*Q $ the co-distribution of ``admissible constraints''
 anihilating ${\cal H}$;
  dually, one has
a decreasing filtration of {\it derived ideals} ending in zero. 

\paragraph{Internal symmetries of NH systems: Noether's theorem.} 
An {\it internal symmetry} occurs whenever a vectorfield $\xi_Q \in {\cal H}$   preserves the Lagrangian. For  natural systems  $\xi_Q $ is a Killing vectorfield for the metric.   Noether's theorem from unconstrained mechanics remains true. The argument (cf. \cite{Arnold2}) goes as follows: denote by
$\phi_{\xi}(s)$ the 1-parameter  group  generated by $\xi$ 
and let
$\phi(s,t) = \phi_{\xi}(s)\cdot q(t)$, so $\phi' = \frac{d}{ds} \phi = \xi_Q (\phi)$.
where $q(t)$ is chosen as a
trajectory of the nonholonomic system. Differentiating with respect to
$s$ the identitly  $L(\phi(s,t),  \frac{d}{dt} \phi(s,t)) = {\rm const.}$,
after a standard integration by parts we get
$ \frac{d}{dt} ( \frac{\partial L}{\partial \dot{q}} \phi') = [L] \phi'\,\, .$
This vanishes precisely when  $\phi' = \xi_Q \in {\cal H }$ so 
$ I_{\xi} :\,= \frac{\partial L}{\partial \dot{q}}\, \cdot \, \xi  = {\rm const.}$

\paragraph{External symmetries:   $G$-Chaplygin  systems.}
  {\it External (or transversal) symmetries} occur when group $G$ acts on $Q$,
 preserving the Lagrangian and the distribution ${\cal H}$, this meaning that $g_* {\cal H}_q = {\cal H}_{gq} $. In the most favorable case one has a principal bundle action
$ G^r \hookrightarrow Q^n \rightarrow S^m,\,\, m+r = n$, 
where ${\cal H}$ forms the horizontal spaces of a connection with 1-form $\phi: TQ \rightarrow Lie(G)$. 
These systems are called   $G-Chaplygin$\footnote{A ``historical''  remark  (by JK). Chaplygin considered   the abelian case.
During a post-doctoral year  in Berkeley, way back in 1982, 
I became interested in NH systems with symmetries.   Alan directed me to two 
wonderful books: \cite{Hertz}   {\it Foundation of Mechanics} and    \cite{NF}.  
In the latter
I learned about (abelian) Chaplygin systems, presented in coordinates.
I said to Alan that I would like to examine non-abelian group symmetries, and Alan immediately made a diagram on his blackboard, and told me:  ``well, then, the constraints are given by   a connection on a pricipal bundle''. This was the starting point of 
\cite{KoillerARMA}.}.

\paragraph{Terminology. } Since \cite{BatesSniatycki}, and \cite{BKMM}, several authors have called attention on these two types of symmetries.  Reduction of internal symmetries was described already in \cite{Sniatycki98}. 
To stress the difference, reduction of external symmetries  
is called here  {\it compression}. The word
{\it reduction} will be used   for internal symmetries.  

\paragraph{LR systems.}     \cite{Veselovs1}, \cite{Veselovs} considered   Lie groups $Q = G$ with {\it left} invariant metrics, with  constraint distributions given by {\it right} translation of ${\cal D} \subset Lie(G)$, i.e., 
the constraints are given by right invariant forms. For a {\it LR-Chaplygin system},   in addition there is a decomposition 
$Lie(G) = Lie(H) \oplus {\cal D}$, where $H$ is a Lie subgroup such that $Ad_{h^{-1}} D =
h^{-1}\,D\,h = D$. 
Therefore   $ H \hookrightarrow G \rightarrow S = G/H$ is a  $H$-Chaplygin system; the base $S$ is the homogeneous space  of cosets $Hg$. \cite{Fedorov1} considered the case where
 $G$ is compact and that $Lie(H)$ is {\it orthogonal} to $D$ with respect to the
 bi-invariant metric\footnote{These conditions are not met in the marble and rubber
 Chaplygin spheres, see section \ref{Chaplyginspheresection};    
 however, the Veselov's result (theorem \ref{Veselovstheorem} below) on invariant volume forms still holds.}.

\paragraph{Compression of G-Chaplygin systems.} From symmetry, it is clear that  the Lagrange-D'Alembert 
equations  {\it compress} to the base $TS$\footnote{The full dynamics can be reconstructed from the compressed solutions,    horizontal lifting  the trajectories via $\phi$, since the admissible paths are horizontal relative to the connection.  This last step  is  not ``just'' a quadrature; in the
non-abelian case,   a {\it path}-ordered integral is in order.  For  $G = SO(3)$,
see \cite{Levi} found an interesting geometric construction.}. 
 In covariant form, the dynamics takes the form 
$[L^{\phi}] = F(s,\dot{s})$, where $L^{\phi}(s,\dot{s}) = L(s,h(\dot{s}))$ is the compressed Lagrangian in $TS$; $h(\dot{s})$ is the horizontal lift to
 any local section and $F$ is a {\it pseudo-gyroscopic force}\footnote{This nonholonomic  force represents, philosophically,
a {\it conceiled force} in the sense of \cite{Hertz}, having a  geometric origin. 
This force vanishes in some
special cases, not necessarily requiring   the constraints being holonomic. 
Equivalently,  the dynamics in $TS$ is   the geodesic spray
of a {\it modified} affine connection. One adds to the induced Levi-Civita connection
in $TS$ a certain tensor $B(X,Y)$.
This NH connection  in general is  non-metric (\cite{KoillerARMA}).}. 
 In order to write $F$ explicitly,
take group-quasicoordinates $(s,\dot{s},g,\dot{\pi})$.
Write $ q = g \sigma(s)$, with $g \in G$ and  a local section  $\sigma(s)$ of $Q \rightarrow S$.
Fix a basis $X_k$ for the Lie algebra, $[X_K,X_L] = \sum \,c^J_{KL}\,X_J$,
$X(\dot{\pi}) = \sum \dot{\pi}_I\,X_I$. Any tangent vector $\dot{q} \in T_{\sigma(s)}Q$ 
can be written as $\dot{q} = d\sigma(s)\cdot \dot{s} +
X(\dot(\pi))\cdot \sigma(s)$.  Horizontal vectors are represented by
$\dot{\pi} = b(s)\cdot \dot{s}$, where $b(s)$ is an $r \times m$ matrix.
The connection 1-form writes as $\phi(\dot{q}) = \dot{\pi} - b(s)\cdot \dot{s}$.  Then
\begin{equation} \label{NHforce}
[L^{\phi}] = F(s,\dot{s})\,\,,\,\,\,  F = \sum_{K=1}^r \,  \left(\frac{\partial L}{\partial \dot{\pi}_k}\right)^*\,
\sum_{j=1}^m \,\left( \frac{b_{Ki}}{\partial q_j} -   \frac{b_{Kj}}{\partial q_i} +
\sum_{U,V=1}^r\, b_{Ui}b_{Vj}\,c_{UV}^K \right)\,\dot{s}_j \,\,\,.
\end{equation}

\subsection{Main results} 
Using the moving frames method we present results on   two  aspects
of    nonholonomic systems. 
\begin{itemize}
\item  Cartan's equivalence, using Cartan's geometric
description of NH systems via affine connections (\cite{Cartan2}). The objective is to find 
all local invariants.
\item Chaplygin systems: compression of external symmetries, reduction
of internal symmetries. The objective is to generalize Chaplygin's ``reducing factor'' method  (\cite{Chaplygin0}), namely, verify if   Hamiltonization is possible (via {\it conformally symplectic structures}). 

\end{itemize}

\paragraph{Results on Cartan's equivalence.}  In section \ref{Cartansection} 
we analyze NH systems under the {\it affine connection} perspective. 
 We pursue  the (local)  classification programme  proposed by \cite{Cartan2}
 using his equivalence method. See \cite{Koiller1}
and \cite{Tavares}, for a rewrite of  Cartan's paper in modern language.  
Cartan's method of equivalence is a powerful method for uncovering and interpreting all
differential invariants and symmetries in a given geometric structure.
 In \cite{Ehlers} NH systems in
 a 3-manifold with a contact distribution were classified.  Here we go  one step further,
   looking at Engel's distribution in 4-manifolds (see definition below). Our results are summarized in Theorem
\ref{structure}. The ``role model'' here is the rolling penny example (no pun intended).
{\it This is the first such study for a distribution that is not strongly nonholonomic}. 
Next in line is studying the famous Cartan's 2-3-5
distribution.  

\paragraph{Results on $G$-Chaplygin systems.}  Instead of using (\ref{NHforce}) in TS, we may
describe the compressed system in $T^*S$ as
 a almost Hamiltonian system\footnote{For details, see   \cite{Koillerromp}, \cite{KoillerRios}. The Hamiltonian compression
for Chaplygin systems was first explored, in the abelian case, by   \cite{Stanchenko}. The non-closed term   was  described  as a semi-basic 2-form,  depending linearly on the fiber coordinate in $T^*S$, but its geometric content
was not indicated there.}
\begin{equation} \label{compressed}
i_X \Omega_{NH} = dH\,,\,\, H = H^{\phi}: T^*S \rightarrow \Re\,,\,\,\Omega_{NH} = \Omega_{\rm can}^{T^*S} + {\rm (J.K)}  \,\,,
\end{equation}
where $H^{\phi}$ is  the Legendre transform of the
compressed Lagrangian.   ${\rm (J.K)}$ is a semi-basic 2-form on $T^*S$ which in general is 
not closed.
  As one may guess, $J$ is the momentum map, and $K$ is  the  
curvature of the connection. Ambiguities cancel, since $J$ is $Ad^*$-equivariant while $K$ 
is $Ad$-equivariant. The construction is independent of the point $q$
 on the fiber over $s$. 

 Under an  $s \in S$ dependent
time reparametrization, 
$ d\tau = f(s)\,dt \,,\,\,$
 several interesting 
  compressed $G$-Chaplygin  systems become Hamiltonian. A  necessary condition is the existence of an
invariant volume (Theorem \ref{Veselovstheorem}), whose density $F$ produces a candidate $ f = F^{1/(m-1)}\,,\, m = {\rm dim}(S)$ for conformal factor. Chaplygin's
``rubber'' ball (vertical rotations forbidden) is, as far we know, a new example, and generalizes the well known Veselova system in $SO(3)$ (Proposition \ref{rubberdensity}).
We describe the obstruction to Hamiltonization as the 2-form
$i_X \,d(f\Omega_{NH})$  (Theorem \ref{criterionHamiltonizable}) and
 we discuss   further reduction by internal symmetries.
 An example of the latter situation is Chaplygin's  ``marble'' (a hard ball with unequal 
inertia coefficients rolling without slipping on the plane). It is non Hamiltonizable 
in $T^*SO(3)$, and our calculations suggest that it is    also non-Hamiltonizable when {\it reduced} to $T^*S^2$ (heorem \ref{marbleconformal}).  Compare  with \cite{Borisovsphere}.

\paragraph{What  does Hamiltonization accomplish.} Why we focus so much on
the question of Hamiltonizability?
The example of the reduced equations for Chaplygin's skate (after a 2-
dimensional euclidean symmetry is removed) shows that changing time
scale in a nonholonomic systems can completely change its character. 
In this example (see e.g. \cite{KoillerARMA}) the fully reduced equations of motion are not Hamiltonian
because every solution is asymptotic in forward and backward
time to a point, which depends on which solution you choose. However,
after rescaling time the fully reduced equations become Hamiltonian,
namely, the harmonic oscillator. However, this Hamiltonian vector
field is incomplete because along one of the coordinate axes the time
rescaling is not defined\footnote{We thank one of the referees for this observation.}. In light of this example, why
is time rescaling interesting? The answer is that it is interesting mostly in the context of integrability,
where no singularities are removed in the phase space. See section \ref{GChap}. 



\section{Nonholonomic geometry: Cartan equivalence} \label{Cartansection}

A  {\it Cartan nonholonomic structure} is a triple $(Q,\g=\langle \hspace{.05in} \cdot \hspace{.05in},\hspace{.05in} \cdot \hspace{.05in}\rangle,\D )$ where $Q$ is an $n$-dimensional manifold  endowed with a Riemannian metric $\g$ and a rank $r$, totally nonholonomic distribution $\D$.  Our motivation for studying such a structure is a free particle moving in $Q$, nonholonomically constrained to $\D$, with kinetic energy $T=\frac{1}{2}\langle \hspace{.05in} \cdot \hspace{.05in},\hspace{.05in} \cdot \hspace{.05in}\rangle$. The nonholonomic geodesic equations are obtained by computing   accelerations   using the Levi-Civita connection associated with $\g$ and orthogonally projecting the result onto $\D$. 
The projected connection is called a {\it nonholonomic connection} (\cite{Lewis}), and was introduced by \cite{Cartan2}. A distribution $\D$ is {\it strongly nonholonomic} if any basis of vectorfields spanning $\D$ on $U\subset Q$, together with their Lie brackets, span the entire tangent space over $U$.  The equivalence problem for nonholonomic geometry was revisited in \cite{Koiller1} and the generalization to arbitrary nonholonomic distributions was discussed. {\it Engel manifolds  provide the simplest example involving distributions that are not strongly nonholonomic}\footnote{Historical remarks. 
\cite{Cartan2} introduced the equivalence problem for nonholonomic geometry and studied the case of manifolds endowed with  {\it strongly nonholonomic distributions}. In his address, Cartan warned against attempts to study other cases because of the ``plus compliqu\'es'' computations involved. In the meantime strides have been made in the equivalence method by Robert Gardner and his students that allow computations to be made at the Lie algebra level rather than at the group level (\cite{Gardner}). This together with symbolic computation packages such as {\it Mathematica{\small \copyright}} 
make equivalence problems tractable in many important cases. See \cite{Gardner}, \cite{Bryant1}, \cite{Montgomery}, \cite{Grossman}, \cite{Ehlers}, \cite{Hughen}, and \cite{Moseley} for some recent applications.}.

The main question we address is the following. Given two nonholonomic structures $(Q,\g,\D )$ and $(\bar{Q},\bar{\g},\bar{\D} )$, is there a (local) diffeomorphism $f:U\subset Q \rightarrow \bar{U} \subset \bar{Q}$ carrying nonholonomic geodesics in $Q$ to nonholonomic geodesics in $\bar{Q}$? In Cartan's approach, this question is recast as an equivalence problem. The nonholonomic structure is encoded into a subbundle of the frame bundle over $Q$, called a $G$-structure. {\it The diffeomorphism $f$ exists if the two corresponding $G$-structures are locally equivalent}. Necessary and sufficient conditions for the $G$-structures to be equivalent are given in terms of differential invariants found using the {\it method of equivalence}. 

 \paragraph{Outline.} Our main example is the equivalence problem for nonholonomic geometry on an Engel manifold. Let $Q$ be a four-dimensional manifold and $\D$ be a rank two distribution. $\D$ is an Engel distribution if and only if,  for any vectorfields $X$ and $Y$  locally spanning $\D$, and some functions $a,b:Q\rightarrow\Re$, the vectorfields $X$, $Y$, $Z=[X,Y]$ , and $W=a[X,Z]+b[Y,Z]$ form a local basis for $TQ$. By an Engel manifold, we mean a four-dimensional manifold endowed with an Engel distribution. We begin by describing the nonholonomic geodesic equations. In the spirit of Cartan's program, we express them in terms of connection one-forms and (co)frames  adapted to the distribution. This formulation is particularly well suited to the problem at hand; the nonholonomic geodesic equations are obtained by writing the ordinary geodesic equations in terms of the Levi-Civita connection one-form and  crossing out terms corresponding to directions complementary to $\D$. We then set up the equivalence problem for nonholonomic geometry and  give   a brief description of the equivalence method as it is applied to our main example. We conclude this section by applying the method of equivalence to the case of nonholonomic geometry on an Engel manifold. 
We derive all differential invariants associated with the nonholonomic structure and show that the symmetry group of such a structure has dimension at most four. 


%
 


\subsection{Nonholonomic geodesics: straightest paths}

\paragraph{Totally nonholonomic distributions.} A distribution $\D$ is a rank $r$ vector subbundle of the tangent bundle $T(Q)$ over $Q$. Let $\D^{1}=\D+[\D,\D ]$ and $\D^{i}=[\D ,\D^{i}]$, and consider the filtration $$\D \subset \D^{1} \subset \cdot \cdot \cdot \D^{i} \subset \cdot \cdot \cdot \subset TQ.$$ 
$\D$ is totally nonholonomic if and only if, for some $k$, $\D^{k}=TQ$ at all points in $Q$. For the present discussion we will assume that rank of each $\D^{i}$ have constant rank over $Q$.
As  a specific example, consider the Engel distribution $\D$ on $\Re^{4}$ with coordinates $(x,y,z,w)$, spanned by $\{ X_{1}=\frac{\partial}{\partial w},X_{2}=\frac{\partial}{\partial x}+w\frac{\partial}{\partial y}+y\frac{\partial}{\partial z}\}$. There are, in fact, local coordinates on any Engel manifold so that the distribution is given by this normal form, see \cite{Montgomery}. Then $\{ X_{1},X_{2},X_{3}=[X_{1},X_{2}]\}$ spans the three-dimensional distribution $\D^{1}$, and $\{ X_{1},X_{2},X_{3},X_{4}=[X_{2},X_{3}]\}$ spans the entire 
$T\Re^4$.

 A path $c:\Re \rightarrow Q$ is {\it horizontal} if $\dot{c}(t) \in \D_{c(t)}$ for all $t$.
 Chow's theorem implies that if $\D$ is totally nonholonomic then any two points in $Q$ can be joined by a horizontal path (see \cite{Montgomery}). At the other extreme, the classical theorem of Frobenius  implies that $\D$ is {\it integrable}, which is to say that $Q$ is foliated by submanifolds whose tangent spaces coincide with $\D$ at each point, if and only if $[X_{i},X_{j}]\in \D$ for all $i$ and $j$ (\cite{Warner}). 

In what follows we will need a description of distributions in terms of differential ideals. Details can be found in \cite{Warner} or \cite{Montgomery}.
Let $\I= \D^{\perp}$ be the ideal in $\Lambda^{*}(Q)$ consisting of the differential forms annihilating $\D$.  If $\D$ is rank $r$, then $\I$ is generated by $n-r$ independent one-forms. The {\it first derived ideal} of $\I$ is the ideal 
 \begin{equation}
 (\I )' := \{\theta \in \I  \hspace{.05in} | \hspace{.05in} d\theta \equiv  0 \hspace{.05in} \mbox{mod}  \hspace{.05in} (\I ) \}.
 \end{equation}

If we set $\I^{(0)}=\I$ and $\I^{(n+1)}=(\I^{(n)})' $ we obtain a decreasing filtration $$\I=\I^{(0)} \supset \I^{(1)} \supset \cdot \cdot \cdot \supset 0.$$ The filtration terminating with the 0 ideal is equivalent to the assumption that the distribution is completely nonholonomic. We note that $I^{(j)}=(\D^{j})^{\perp}$ for $j=1$, but this is not true in general for $j>1$ (see \cite{Montgomery}). At the other extreme,  the differential ideal version of the Frobenius theorem implies that $\D$ is integrable if and only if $(\I)'\subset \I$ (\cite{Warner}).

For the Engel example, the one forms $\eta^{1}=dy-wdx$ and $\eta^{2}=dz-ydx$ generate the ideal $\I$. Notice that $d\eta^{2}=\eta^{1}\wedge dx$ so $\eta^{2}\in \I^{(1)}$ but $d\eta^{1}$ cannot be written in terms of $\eta^{1}$ or $\eta^{2}$ therefore $\eta^{1}\notin \I^{(1)}$.

\paragraph{The nonholonomic geodesic equations.} There are two different geometries commonly defined on a nonholonomic structure $(Q,\g=\langle \hspace{.05in} \cdot \hspace{.05in},\hspace{.05in} \cdot \hspace{.05in}\rangle,\D )$:   {\it subriemannian geometry and nonholonomic geometry}. In subriemannian geometry one is interested in {\it shortest paths.} The length of a path $c:[a,b]\rightarrow Q$ joining points $x$ and $y$ is $\ell(c)=\int \sqrt{\langle \dot{c},\dot{c}\rangle}dt$. The distance from $x$ to $y$ is $d(x,y)=\mbox{inf}(\ell (c))$ taken over all horizontal paths joining $x$ to $y$.
In nonholonomic geometry one is interested in {\it straightest paths}, which are   solutions to the nonholonomic geodesic equations. \cite{Hertz} was the first to notice that shortest $\neq $ straightest unless the constraints are holonomic\footnote{The terminology {\it straightest path} for a nonholonomic geodesic was in fact coined by Hertz himself.}.


The nonholonomic geodesic equations are obtained by computing the acceleration of a horizontal path $c:\Re \rightarrow Q$ using the Levi-Civita connection  associated with $\g$ and orthogonally projecting the result onto $\D$. It is convenient to adopt the following indicial conventions:
\begin{eqnarray}
1\leq & I, J, K & \leq n \nonumber \\
1\leq & i, j, k & \leq r \hspace{.1in} (=\mbox{rank} (\D ))  \\
r+1 \leq & \nu & \leq n.  \nonumber
\end{eqnarray}

Let $e=\{e_{I} \}$ be a local orthonormal frame for which the $e_{i}$ span $\D$, and let $\eta =\{ \eta_{I} \}$ be the dual coframe defined by $\eta_{I} (e_{J} )=\delta_{IJ}$, the Kronecker delta function. We note that the $\eta^{\nu}$  annihilate $\D$ and the metric, restricted to $\D$ is $g_{|\D} =\eta^{1}\otimes \eta^{1}+ \cdot \cdot \cdot +\eta^{r}\otimes \eta^{r}$. The Levi-Civita connection can be expressed in terms of local one-forms $\omega_{IJ}=-\omega_{JI}$ satisfying Cartan's structure equation  $d\eta=-\omega \wedge \eta$ (\cite{Hicks}). 

 A horizontal path $c: \Re \rightarrow M$ is a {\it nonholonomic geodesic} if it satisfies the nonholonomic geodesic equations 
\begin{equation}
\left[ \frac{d}{dt} (v_{i})+\sum_{j}v_{j} \omega_{ij} (\dot{c} ) \right] e_{i}=0, 
\end{equation}
 where $1\leq i,j,\leq r$ and $v_{i}=\eta^{i}(\dot{c})$ are the quasivelocities.

\paragraph{Example: the vertical rolling penny.}
 A standard example of a mechanical system modeled by a nonholonomic Engel system is that of a coin rolling without sliping on the euclidean plane. Consider a coin of radius $a$ rolling vertically on the $xy$-plane. 
The location of the coin is represented by the coordinates $(x,y,\theta,\phi)$. The point of contact of the coin with the plane is $(x,y)$,  the angle made by the coin with respect to the positive $x$-axis is $\theta$, and the angle made by the point of contact, the center of the coin, and a point marked on the outer edge of the coin is $\phi$. The state space can be identified with the Lie group $SE(2)\times SO(2)$ where the first factor is
the group of Euclidean motions locally parametrized by $x$, $y$ and $\theta$. The mass of the coin is $m$, the moment of inertia in the $\theta$ direction is $J$ and the moment of inertia in the $\phi$ direction is $I$. The kinetic energy, which defines a Riemannian metric on the state space, is 
\begin{equation}
T = \frac{m}{2}(dx\otimes dx+dy\otimes dy)+\frac{J}{2}d\theta \otimes d\theta +\frac{I}{2}d\phi \otimes d\phi .
\end{equation}
The penny rolls without slipping giving rise to the constraints 
\begin{equation}
\dot{x}=(a \cos \theta )\, \dot{\phi} \,\,\,,\,\,\,
\dot{y}=(a \sin \theta ) \, \dot{\phi} \,\,\,\,.
\end{equation}
Consider the orthonormal frame $(X_{1},X_{2},X_{3},X_{4})$ where
\begin{eqnarray}
& X_{1}:=\sqrt{\frac{2}{ma^{2}+I}}\left( a\cos \theta \frac{\partial}{\partial x} +a\sin \theta \frac{\partial}{\partial y}+\frac{\partial}{d \phi}\right) \,\,\,,\,\,\,  X_{2}:=\sqrt{\frac{2}{J}}\frac{\partial}{\partial \theta}  & \\
& X_{3}:=\sqrt{\frac{2}{m}}\left( -\sin \theta \frac{\partial}{\partial x}+\cos \theta \frac{\partial}{\partial y}\right)  \,\,\,,\,\,\,
X_{4}:=\sqrt{\frac{2}{m}}\left(\cos \theta \frac{\partial}{\partial x}+\sin \theta \frac{\partial}{\partial y} \right).
\nonumber
\end{eqnarray}
Note that the constraint subspace ${\cal H}=\mbox{span}\{ X_{1},X_{2}\}$, and ${\cal H}^{(1)}=\mbox{span}\{ X_{1},X_{2},X_{3}\}$. 
The dual coframe is $(\eta^{1},\eta^{2},\eta^{3},\eta^{4})$ where
\begin{eqnarray}
& \eta^{1}:= \sqrt{\frac{ma^{2}+I}{2}}d\phi \,\,\,,\,\,\,
\eta^{2}:= \sqrt{\frac{J}{2}}d\theta  &  \label{eq:oncoframe} \\
& \eta^{3} := \sqrt{\frac{m}{2}}(-\sin \theta dx+\cos \theta dy)  \,\,\,,\,\,\, \eta^{4} := \sqrt{\frac{m}{2}}(\cos \theta dx+\sin \theta dy-d\phi). & \nonumber  
\end{eqnarray}

To compute the Levi Civita connection form we determine $\omega=[\omega_{IJ}]$ such that $\omega_{IJ}=-\omega_{JI}$ and $d\eta = -\omega \wedge \eta.$ Using simple linear algebra we find
\begin{equation}
{\bf \omega}= 
 \left(
\begin{array}{cccc} 
 0&\frac{1}{\sqrt{2}}\sqrt{\frac{m}{J(ma^{2}+I)}}\hspace{.05in} \eta^{3} &\frac{1}{\sqrt{2}}\sqrt{\frac{m}{J(ma^{2}+I)}}\hspace{.05in} \eta^{2}&0 \\ 
&&&\\
-\frac{1}{\sqrt{2}}\sqrt{\frac{m}{J(ma^{2}+I)}}\hspace{.05in} \eta^{3}&0&-\frac{1}{\sqrt{2}}\sqrt{\frac{m}{J(ma^{2}+I)}}\hspace{.05in} \eta^{1}&0\\
&&&\\
-\frac{1}{\sqrt{2}}\sqrt{\frac{m}{J(ma^{2}+I)}}\hspace{.05in} \eta^{2}&\frac{1}{\sqrt{2}}\sqrt{\frac{m}{J(ma^{2}+I)}}\hspace{.05in} \eta^{1}&0&-\frac{\sqrt{2}}{\sqrt{J}}\eta^{2}  \\
&&&\\
0&0&\frac{\sqrt{2}}{\sqrt{J}}\eta^{2}&0  
\end{array} \right) 
\end{equation}
so in particular $$ \omega_{12}=-\omega_{21}=\frac{1}{2}\sqrt{\frac{m}{J(ma^2+I)}} \hspace{.1in} \eta^{3}.$$
\bigskip

 Let $c:\Re\rightarrow Q$ be a nonholonomic geodesic given by $\dot{c}(t)=v_{1}(t)X_{1}+v_{2}(t)X_{2}.$ From the structure equations we see immediately that $ \omega_{12}(\dot{c}(t))=-\omega_{21}(\dot{c}(t))=0$  and the nonholonomic  geodesic  equations  reduce to $\frac{d}{dt}(v_{1})=\frac{d}{dt}(v_{2})=0.$ The nonholonomic geodesics are solutions to $(\dot{x},\dot{y},\dot{\phi},\dot{\theta})=AX_{1}+BX_{2}.$ In particular,
\begin{equation}
\dot{x}=\frac{\sqrt{2}Aa\cos \theta(t)}{\sqrt{ma^{2}+I}}, \hspace{.2in}  \dot{y}=\frac{\sqrt{2}Aa\sin \theta(t)}{\sqrt{ma^{2}+I}}, \hspace{.2in}
\dot{\phi}=\frac{\sqrt{2}A}{\sqrt{ma^{2}+I}}, \hspace{.2in} \dot{\theta}=\frac{\sqrt{2}B}{\sqrt{J}}.
\end{equation}
The trajectories are spinning in place ($A=0$), rolling along a line ($B=0$), or circles ($A,B\neq 0$).

\subsection{Equivalence problem of nonholonomic geometry}


Cartan's method of equivalence starts by encoding a geometric structure in terms of a subbundle of the coframe bundle called a $G$-structure. We begin this section by describing the $G$-structure for nonholonomic geometry\footnote{This $G$-structure was first presented by Cartan in his 1928 address to the International Congress of Mathematicians (\cite{Cartan2})}. We then give a brief outline of some of the main ideas behind the method of equivalence as it is applied in our example of nonholonomic geometry on an Engel manifold. Details on the method of equivalence can be found in \cite{Gardner}, \cite{Montgomery}, or \cite{Bryant1}. We  then derive the local invariants associated with a nonholonomic structure on a 4-dimensional manifold endowed with an Engel distribution.

\paragraph{Initial $G$-structure for nonholonomic geometry.} 
A {\it coframe} $\eta(x)$ at $x  \in Q^n$ is a basis for the cotangent space $T_{x}^{*}(Q)$. Alternatively, we can regard a coframe as a linear isomorphism $\eta (x):T_{x}(Q)\rightarrow \Re^{n}$ where $\Re^{n}$ is represented by column vectors. A coframe can then be multiplied by a matrix on the left in the usual way. The set of all coframes at $x$ is denoted $F_{x}^{*}(Q)$ and has the projection mapping $\pi :F_{x}^{*}(Q) \mapsto x$. The coframe bundle $F^{*}(Q)$ is the union of the $F_{x}^{*}(Q)$ as $x$ varies over $Q$. A coframe is a smooth (local) section $\eta :Q\rightarrow F^{*}(Q)$ and is represented by a column vector of one-forms  $(\eta^{1},...,\eta^{n})^{\mbox{tr}}$, where $\,$ ``tr'' $\,$ indicates transpose. $F^{*}(Q)$ is a right $Gl(n)$-bundle with action $R_{g}\eta =g^{-1}\eta$ where $g$ is a matrix in $Gl(n)$. 
\smallskip
Let $G$ be a matrix subgroup of $Gl(n)$. A {\it $G$-structure} is  a $G$-subbundle of $F^{*}(Q)$.
 We now describe the $G$-structure encoding the nonholonomic geometry associated with a nonholonomic structure $( Q, \g ,\D )$.
Given  a nonholonomic structure $(Q, \g ,\D )$ we can choose an orthonormal coframe $\eta=(\eta^{i},\eta^{\nu})^{\mbox{tr}}$ on $U\subset Q$ so that the $\eta^{\nu}$ annihilate $\D$ and use this coframe to write down the nonholonomic geodesic equations as described above. On the other hand, given a coframe $\bar{\eta}=(\bar{\eta}^{i},\bar{\eta}^{\nu})^{\mbox{tr}}$ on $Q$ we can construct a nonholonomic structure $(Q, \bar{\g}=\sum \bar{\eta}^{i} \otimes \bar{\eta}^{i}+\bar{\eta}^{\nu} \otimes \bar{\eta}^{\nu}, \bar{\D})$ where $\bar{\D}$ is annihilated by the $\bar{\eta}^{\nu}$. How is $\bar{\eta}$  related to $\eta$ if it is to lead to the same nonholonomic geodesic equations as $\eta$?
  In order to preserve $\D$ we must have  $\eta^{\nu}-\bar{\eta}^{\nu}=0$ (mod $I$). In matrix notation, any modified coframe $\bar{\eta}$ must be related to $\eta$  by
\begin{equation}   \label{eq:distribution}
\left(
\begin{array}{c} 
\bar{\eta}^{i}\\ 
\bar{\eta}^{\nu}
\end{array} \right)
 =
\left(
\begin{array}{cc} 
A &b\\ 
0 & a 
\end{array} \right) 
\left(
\begin{array}{c} 
\eta^{i}\\ 
\eta^{\nu}
\end{array} \right) . 
\end{equation}
where $A\in Gl(r)$, $a\in Gl(n-r)$, and $b\in M(k,n-r)$.  If we were studying the geometry of distributions there would be no further restrictions. In order to preserve the metric restricted to $\D$, we must further insist that $A\in O(r)$. We would then have the starting point for the study of subriemannian geometry  (see \cite{Montgomery}, \cite{Hughen}, or \cite{Moseley}). 

\bigskip

{\it It is   important to observe that  that in nonholonomic geometry we need  the full metric and not just its restriction to $\D$ (as in subriemannian geometry) to obtain the equations of motion. }
\cite{Cartan2} showed that {\it in order to preserve the nonholonomic geodesic equations, we can only add  covectors that are in the  first derived ideal to the} $\eta^{i}$. 

\bigskip

Since this fact is central to our analysis, we sketch the argument here (see  \cite{Koiller1} for details).  Suppose $\bar{\eta}=g\eta$ with connection one-form defined by $d\bar{\eta}=-\bar{\omega}\wedge \bar{\eta}$.
For simplicity assume that $A=id$, then $\eta^{j}\equiv \bar{\eta}^{j} \hspace{.05in} (\mbox{mod} \hspace{.05in}  \I)$. The geodesic equations are preserved if and only if $\omega_{ij}(T)=\bar{\omega}_{ij}(T)$ for all $T\in \D$, in other words $\omega_{ij}\equiv \bar{\omega}_{ij} \hspace{.05in} (\mbox{mod} \hspace{.05in} \I)$.  Note also that $\bar{\eta}^{\nu} \equiv 0 \hspace{.05in} (\mbox{mod} \hspace{.05in} \I)$.
Subtracting the structure equations for $d\eta^{i}$ and $d\bar{\eta}^{i}$ we get
$$
d\eta^{i}-d\bar{\eta}^{i} = -\omega_{ij} \wedge \eta^{j}-\omega_{i\nu} \wedge \eta^{\nu}+\bar{\omega}_{ij} \wedge \bar{\eta}^{j}+\bar{\omega}_{i\nu} \wedge \bar{\eta}^{\nu} 
 \equiv \hspace{.05in} 0 \hspace{.05in} (\mbox{mod} \hspace{.1in} \I).  
$$
Now $\bar{\eta}^{i}=\eta^{i}+b_{i\nu} \eta^{\nu}$ so we also have
$$
d\eta^{i}-d\bar{\eta}^{i} = d\eta^{i}-(d\eta^{i}+db_{i\nu} \eta^{\nu}+b_{i\nu} d\eta^{\nu}) 
\equiv  -b_{i\nu} d\eta^{\nu} \hspace{.1in} (\mbox{mod} \hspace{.1in} \I) 
$$
Therefore $b_{i\nu} d\eta^{\nu}\equiv 0 \hspace{.1in} (\mbox{mod} \hspace{.1in} \I)$ or equivalently $b_{i\nu} \eta^{\nu} \in I^{(1)}$. This completes the argument.

We further subdivide our indicial notation:  let
$$
r+1\leq  \phi  \leq s \hspace{.1in} \mbox{(= rank $\D^{1}$)} \,\,\,\,,\,\,\,\, s+1\leq  \Phi  \leq n.
$$

\paragraph{Adapted coframes.} A covector 
 $\,\,\, \eta=(\eta^{i},\eta^{\phi},\eta^{\Phi})^{\mbox{tr}})\,\,\,$ arranged so that
\begin{enumerate} 
\item The $\eta^{\phi}$ and $\eta^{\Phi}$ generate $I$,
\item  $ds^{2}|_{\D}=\sum \eta^{i}\otimes \eta^{i}$,
\item The $\eta^{\Phi}$ generate the first derived ideal $I^{(1)}$,
\end{enumerate}
is said to be {\it adapted to the nonholonomic structure}.
In matrix notation, the most general change of coframes that preserves the nonholonomic geodesic equations is of the form $\bar{\eta}=g\eta$ where
\begin{equation}
g=\left(
\begin{array}{ccc} 
A &0&b\\ 
0 & a_{1}&a_{2} \\
 0 & 0 &a_{3}  \label{eq:G00}
\end{array} \right) 
\end{equation}
with $A\in O(k)$, $b\in M(n-s,k)$, $a_{1}\in Gl(s-k)$, $a_{2}\in Gl(n-s,s-k)$, and $a_{3}\in Gl(n-s)$.
The set of all such block matrices form a matrix subgroup of $Gl(n)$ which we shall denote $G_{0}$.

\smallskip

  The {\it initial $G$-structure} for nonholonomic geometry on $(Q,ds^{2},\D)$  is a subbundle $B_{0}(Q) \subset F^{*}(Q)$ (or simply $B_{0}$ if there is no risk of confusion) with structure group $G_{0}$ defined above. 
All local sections of $B_{0}(Q)$ lead to the same nonholonomic geodesic equations. In this way, the initial $G$-structure $B_{0}(Q)$ completely characterizes the nonholonomic geometry.

Two $G$-structures, $B(Q)\stackrel{\pi_{Q}}{\rightarrow}Q$ and $B(N)\stackrel{\pi_{N}}{\rightarrow}N$, are said to be {\it equivalent} if there is a diffeomorphism $f:Q\rightarrow N$ for which $f_{1}(B(Q))=B(N)$ where $f_{1}$ is the induced bundle map. (If we think of $b\in B(Q)$ as a linear isomorphism $b:T_{\pi_{Q} (b)}Q \rightarrow \Re^{n}$ then $f_{1}(b)=b\circ (f_{*})^{-1}$ where $f_{*}$ is the differential of $f$.) Our original question of whether there is a local diffeomorphism that carries nonholonomic geodesics to nonholonomic geodesics can be answered by determining whether the associated $G$-structures are locally equivalent. 

\subsection{A tutorial on the method of equivalence}
  Necessary and sufficient conditions for the equivalence between $G$-structures are given in terms of {\it differential invariants} which are derived using the method of equivalence. In this section we briefly describe some of the main ideas behind the method of equivalence as it is applied in our example. Details and other facets of the method together with many examples can be found in the excellent text by Robert Gardner (\cite{Gardner}).  
One of the principal objects used in the method of equivalence is the {\it tautological one-form}.  Let $B(Q)\stackrel{\pi}{\rightarrow}Q$ be a $G$-structure with structure group $G$ whose Lie Algebra is $Lie(G$). The tautological one-form  $\Omega$ on $B(Q)$  is an $\Re^{n}$-valued one-form defined as follows. Let $\eta :U\subset Q\rightarrow B(Q)$ be a local section of $B(Q)$ and consider the  inverse trivialization $U\times G_{0}\rightarrow B(Q)$ defined by $(x,g) \rightarrow g^{-1} \eta(x)$. Relative to this section, the tautological one-form is defined by
\begin{equation}
\Omega(b)=g^{-1}(\pi^{*}\eta ) \label{eq:tautological}
\end{equation}
 where $b=g^{-1} \eta$.  From (\ref{eq:tautological}) one can verify that the tautological one-form is semi-basic (i.e. $\Omega (v)=0$ for all $v\in \hspace{.02in} ker(\pi_{*})$), has the reproducing property $\bar{\eta}^{*}\Omega =\bar{\eta}$ where $\bar{\eta}$ is any local section of $B(Q)$, and is equivariant: $R_{g}^{*}\Omega = g^{-1}\Omega$. The components of the tautological one-form provide a partial coframing for $B(Q)$ and form a basis for the semi-basic forms on $B(Q)$. 

The following proposition  reduces the problem of finding an equivalence between $G$-structures to finding a smooth map that preserves the tautological one-form. (See \cite{Gardner} or \cite{Bryant1} for a proof.)
\begin{proposition}
Let $B(Q)$ and $B(N)$ be two $G$-structures with corresponding tautological one-forms $\Omega_{Q}$ and $\Omega_{N}$, and let $F:B(Q)\rightarrow B(N)$ be a smooth map. If $G$ is a connected and $F^{*}(\Omega_{N})=\Omega_{Q}$ then there exists a local diffeomorphism $f:Q\rightarrow N$ for which $F=f_{1}$, i.e. the two $G$-structures are equivalent.
\end{proposition}
To find the map $F$ in this proposition we would like to apply {\it Cartan's technique of the graph} (cf. \cite{Warner} p. 75): if we could find an integral manifold $\Sigma \subset B(Q)\times B(N)$ of the one-form $\theta = \Omega_{Q}-\Omega_{N}$ that projects diffeomorphically onto each factor, then $\Sigma$ would be the graph of a function $h:Q\rightarrow N$ for which $h_{1}^{*}\Omega_{N}=\Omega_{Q}$. By the above proposition the $G$-structures would then be equivalent. We generally cannot apply this idea directly because $\Omega_{Q}$ and $\Omega_{N}$ do not provide full coframes on $B(Q)$ and $B(N)$ as is required in the technique of the graph. In the example of nonholonomic geometry on Engel manifolds, and indeed in many important examples (see \cite{Gardner}, \cite{Hughen}, \cite{Moseley}, \cite{Montgomery}, \cite{Ehlers}), application of the method of equivalence leads to a new $G$-structure called an $e$-structure. An $e$-structure is a $G$-structure endowed with a canonical coframe.

Differentiating both sides of (\ref{eq:tautological}) one can verify that  $d\Omega$ satisfies the structure equation 
\begin{equation}
d\Omega =-\alpha \wedge \Omega+T \label{eq:torsion}
\end{equation}
where $T$ is a semi-basic two-form on $B(Q)$ and $\alpha$ is a called a {\it pseudoconnection}: a $Lie(G)$-valued one-form on $B(Q)$ that agrees with the Mauer-Cartan form on vertical 
vectorfields. Here, $Lie(G)$ is the Lie Algebra of $G$. Summarizing,
 \begin{equation}
 {\rm Pseudoconnection:}\,\,\,\, \alpha = g^{-1}dg + {\rm semibasic} \,\,Lie(G) {\rm -valued \,\,one \,\,form}.
 \end{equation}

The components of the pseudoconnection together with the tautological one-form do provide a full coframe on the $G$-structure, but unlike the tautological one-form, the pseudoconnection is not canonically defined. {\it Understanding how changes in the pseudoconnection affect the torsion is at the heart of the method of equivalence}. 

For any $G$-structure, that part of the torsion that is left unchanged under all possible changes of pseudoconnection is known as the {\it intrinsic torsion}. The intrinsic torsion is the only first order differential invariant of the $G$-structure (\cite{Gardner}). As an example, the intrinsic torsion for the $G$-structure $B$ of a general distribution (equation \ref{eq:distribution})  is the dual curvature of the distribution (\cite{Cartan3}, see also \cite{Montgomery}). In the case of a rank two distribution on a four dimensional manifold, the structure equations for the tautological one-form $\Omega$ are
\begin{eqnarray}
d\left( \begin{array}{l} 
\Omega_{1} \\ 
\Omega_{2}\\
\Omega_{3}\\
\Omega_{4}
\end{array} \right)
=
-\left(
\begin{array}{llll} 
{\cal A}_{11} & {\cal A}_{12} & \beta_{13} &\beta_{14}\\ 
{\cal A}_{21} & {\cal A}_{22} & \beta_{23} & \beta_{24}\\
  0& 0  & \alpha_{33} & \alpha_{34}\\
  0& 0  & \alpha_{34} & \alpha_{44}
\end{array} \right)
\wedge
\left( \begin{array}{l} 
\Omega^{1} \\ 
\Omega^{2}\\
\Omega^{3}\\
\Omega^{4}
\end{array} \right)
+
\left( \begin{array}{c} 
T^{1} \\ 
T^{2}\\
T^{3}\\
T^{4}
\end{array} \right)
\end{eqnarray}
where $T^{I}=\sum_{J<K}T^{I}_{JK}\Omega^{J}\wedge \Omega^{K}$ with $T^{I}_{JK} :B\rightarrow \Re$. The intrinsic torsion consists of the terms $T^{3}_{12}\Omega^{1}\wedge \Omega^{2}$ and $T^{4}_{12}\Omega^{1}\wedge \Omega^{2}$. Note that the distribution is integrable if and only if  $T^{3}_{12}=T^{4}_{12}=0$.

\paragraph{Reduction and prolongation.}
There are two major steps in the equivalence method: 
{\it prolongation} and {\it reduction} (see \cite{Gardner} or \cite{Montgomery}). In the case of nonholonomic geometry on an Engel manifold a sequence of reductions lead to an $e$-structure. A brief outline of the reduction procedure is as follows. The first step involves writing out the structure equations for the tautological one-form $\Omega$. A semi-basic $Lie(G)$-valued one-form is added to the pseudoconnection to make the torsion as simple as possible. \cite{Gardner} calls this step {\it absorption of torsion.}  The action of $G$ on the torsion is deduced by differentiating both sides of the identity  $R_{g}^{*}(\Omega)=g^{-1}\Omega$. The action of $G$ is used to simplify part of the torsion. The isotropy subgroup of that choice of simplified torsion is then the structure group of the reduced $G$-structure. In the case of nonholonomic geometry on an Engel manifold this procedure is repeated until an $e$-structure is obtained.

Suppose that $\Omega$ is the canonical coframing on the resulting a manifold $B$. The $\Omega^{i}$ form a basis for the one-forms on $B$ so we can write 
\begin{equation}
d\Omega^{I}=\sum_{J<K}c_{JK}^{I}\Omega^{J}\wedge \Omega^{K}\,\,.
\end{equation} Relationships between the $c_{JK}^{I}$ are found by differentiating this equation. The resulting 
torsion functions provide the ``complete invariants'' for the geometric structure (see \cite{Gardner} p.59, \cite{Bryant1} pp.9-10, or \cite{Cartan1}). 

Many important examples have {\it integrable $e$-structures}. An $e$-structure is integrable if the $c_{JK}^{I}$ are constant (\cite{Gardner}). In this case we can apply the following result 
from \cite{Montgomery}:
\begin{lemma} \label{framing}
Let $B$ be an $n$-dimensional manifold endowed with a coframing $\Omega$. Then the (local) group $G$ of diffeomorphisms of $B$ that preserves this coframing is a finite-dimensional (local) Lie group of dimension at most $n$. The bound $n$ is achieved if and only if the $e$-structure is integrable. In this case the $c_{JK}^{I}$ are the structure constants of $G$, $G$ acts freely and transitively on $B$, and the coframe can be identified with the left invariant one-forms on $G$.
\end{lemma}
The Jacobi identities are found by differentiating $d\Omega^{i}=\sum_{J<K}c_{JK}^{I}\Omega^{J}\wedge \Omega^{K}$. Lie's third fundamental theorem then implies that we can, at least in principle, reconstruct the group $G$ using the structure constants. In some circumstances one can also conclude that $B$ itself is a Lie group (see \cite{Gardner} p.72).

\subsection{The nonholonomic geometry of an Engel manifold.}
 The initial $G$ structure for nonholonomic geometry on $\{ Q,\g,  \D  \}$ where $\D$ is an Engel distribution on a four-dimensional manifold $M$  is the subbundle $B_{0}\subset F^{*}(Q)$ with structure group $G_{0}$ consisting of matrices of the form 
\begin{equation}
\left(
\begin{array}{cccc} 
A_{11} & A_{12} & 0 &B_{14}\\ 
A_{21} & A_{22} & 0 & B_{24}\\
  0& 0  & a_{33} & a_{34}\\
  0& 0  & 0 & a_{44}
\end{array} \right)  \label{eq:G0}
\end{equation}
where $A=[A_{IJ}]\in O(2)$, $a_{33}a_{44}\neq 0$, and $B_{14}$ and $B_{24}$ are arbitrary.

Let $\Omega = (\Omega^{1},\Omega^{2},\Omega^{3},\Omega^{4})^{tr}$ be the tautological one-form on $B_{0}$. The structure equations are
\begin{eqnarray}
d\left( \begin{array}{l} 
\Omega_{1} \\ 
\Omega_{2}\\
\Omega_{3}\\
\Omega_{4}
\end{array} \right)
=
-\left(
\begin{array}{llll} 
0 & \gamma & 0 &\beta_{14}\\ 
-\gamma & 0 & 0 & \beta_{24}\\
  0& 0  & \alpha_{33} & \alpha_{34}\\
  0& 0  & 0 & \alpha_{44}
\end{array} \right)
\wedge
\left( \begin{array}{l} 
\Omega^{1} \\ 
\Omega^{2}\\
\Omega^{3}\\
\Omega^{4}
\end{array} \right)
+
\left( \begin{array}{c} 
T^{1}_{13}\Omega^{1} \wedge \Omega^{3}+T^{1}_{23}\Omega^{2} \wedge \Omega^{3} \\ 
T^{2}_{13}\Omega^{1} \wedge \Omega^{3}+T^{2}_{23}\Omega^{2} \wedge \Omega^{3}\\
T^{3}_{12}\Omega^{1} \wedge \Omega^{2}\\
T^{4}_{13}\Omega^{1} \wedge \Omega^{3} +T^{4}_{23}\Omega^{2} \wedge \Omega^{3}
\end{array} \right)
\end{eqnarray}
where we have chosen the pseudoconnection so that the remaining $T^{i}_{jk}$ are zero. $\Omega^{4} \in I^{(1)}$ so $d\Omega^{4}=0$ mod $(\Omega^{3},\Omega^{4})$ and we must therefore have $T^{4}_{12}=0$. Also, $\Omega^{3} \notin I^{(1)}$ so $d\Omega^{3}\neq 0$ mod $(\Omega^{3},\Omega^{4})$ therefore the torsion function $T^{3}_{12}$  cannot equal zero.
The pseudo-connection for this choice of torsion is not unique. We can, for instance, add  arbitrary multiples of $\Omega^{4}$ to the $\beta_{i4}$ and $\alpha_{i4}$. 

Following Cartan's prescription, we investigate the induced action of $G_{0}$ on the torsion. Let $g\in G_{0}$. To simplify notation,  functions and forms pulled back by $R_{g}$ will be indicated by a hat so, for instance, $R_{g}^{*} \Omega =\hat{\Omega} =(\hO^{1},\hO^{2},\hO^{3},\hO^{4})^{\mbox{tr}}$ and $R_{g}^{*}(T_{ij}^{k})=\hat{T}_{ij}^{k}$. We have
\begin{equation}
\left( \begin{array}{l} 
\hO^{1} \\ 
\hO^{2}\\
\hO^{3}\\
\hO^{4}
\end{array} \right)=
\left( \begin{array}{c} 
\# \\ 
\#  \\
\mbox{det}( a^{-1})(a_{44}\Omega^{3}-a_{34}\Omega^{4})\\
\mbox{det}(a^{-1})(a_{33}\Omega^{4})
\end{array} \right) \,\,\,\,.
\end{equation}

To determine the induced action of $G_{0}$ on the torsion we differentiate both sides of the identity $R_{g}^{*} \Omega^{3} =\hat{\Omega}^{3}$. For $\Omega^{3}$ we compute
\begin{eqnarray}
R_{g}^{*}(d\Omega^{3})&=&\hA_{33}\wedge\hO^{3}+\hA_{34}\wedge\hO^{4}+\hT_{12}^{3} \hO^{1}\wedge \hO^{2} \nonumber \\
&=&\mbox{det} (A^{-1})\hT^{3}_{12}\Omega^{1}\wedge \Omega^{2} \hspace{.3in} (\mbox{mod}\hspace{.1in} \Omega^{3}, \Omega^{4}) \nonumber
\end{eqnarray}
and
\begin{eqnarray}
d\hO^{3}&=&\mbox{det} (a^{-1})(a_{44}d\Omega^{3}-a_{34}d\Omega^{4}) \hspace{.3in} (\mbox{mod}\hspace{.1in} \Omega^{3}, \Omega^{4}) \nonumber \\
&=& \mbox{det} (a^{-1})(a_{44}T^{3}_{12}\Omega^{1}\wedge \Omega^{2})\hspace{.3in} (\mbox{mod}\hspace{.1in} \Omega^{3}, \Omega^{4}) \,\,\,.   \nonumber
\end{eqnarray}
The induced action of $G_{0}$ on $T^{3}_{12}$ is therefore 
\begin{equation}
R_{g}^{*}(T_{12}^{3}) = \frac{\mbox{det}(A)}{a_{33}}T_{12}^{3}
\end{equation}
Since $T_{12}^{3}\neq 0$ we can force it to equal 1 using the action of $G_{0}$. The stabilizer subgroup $G_{1}$ for this choice of torsion consists of matrices of the form (\ref{eq:G0}) with $a_{33}=\epsilon$ where $\epsilon =\mbox{det}(A)$. Note that $T_{12}^{3}\Omega^{1}\wedge \Omega^{2}$ is the (normalized) dual curvature of the distribution.

The structure equations for the $G_{1}$-structure $B_{1}$ are
\begin{eqnarray}
d\left( \begin{array}{l} 
\Omega_{1} \\ 
\Omega_{2}\\
\Omega_{3}\\
\Omega_{4}
\end{array} \right)
=
-\left(
\begin{array}{llll} 
0 & \gamma & 0 &\beta_{14}\\ 
-\gamma & 0 & 0 & \beta_{24}\\
  0& 0  & 0 & \alpha_{34}\\
  0& 0  & 0 & \alpha_{44}
\end{array} \right)
\wedge
\left( \begin{array}{l} 
\Omega^{1} \\ 
\Omega^{2}\\
\Omega^{3}\\
\Omega^{4}
\end{array} \right)
+
\left( \begin{array}{c} 
T^{1}_{13}\Omega^{1} \wedge \Omega^{3}+T^{1}_{23}\Omega^{2} \wedge \Omega^{3} \\ 
T^{2}_{13}\Omega^{1} \wedge \Omega^{3}+T^{2}_{23}\Omega^{2} \wedge \Omega^{3}\\
T_{13}^{3}\Omega^{1}\wedge \Omega^{3}+T_{23}^{3}\Omega^{2}\wedge \Omega^{3}+\Omega^{1} \wedge \Omega^{2}\\
T^{4}_{13}\Omega^{1} \wedge \Omega^{3} +T^{4}_{23}\Omega^{2} \wedge \Omega^{3}
\end{array} \right) \,\,.
\end{eqnarray}
Let $g\in G_{1}$. We write the inverse of $g$ as
\begin{equation}
g^{-1}=\left(
\begin{array}{cccc} 
A_{11} & A_{21} & 0 &\bB_{14}\\ 
A_{12} & A_{22} & 0 & \bB_{24}\\
  0& 0  & \ba_{33} & \ba_{34}\\
  0& 0  & 0 & \ba_{44}
\end{array} \right)  
\end{equation}
so in particular $\ba_{33}=\epsilon$, $\ba_{34}=-\epsilon a_{34}(a_{44})^{-1}$, and $\ba_{44}=(a_{44})^{-1}$. We have
\begin{equation}
R_{g}^{*}\Omega =\left( \begin{array}{l} 
\hO^{1} \\ 
\hO^{2}\\
\hO^{3}\\
\hO^{4}
\end{array} \right)=
\left( \begin{array}{c} 
A_{11}\Omega^{1}+A_{21}\Omega^{2}+\bB_{14}\Omega^{4}\\ 
A_{12}\Omega^{1}+A_{22}\Omega^{2}+\bB_{24}\Omega^{4}  \\
\ba_{33}\Omega^{3}+\ba_{34}\Omega^{4}\\
\ba_{44}\Omega^{4}
\end{array} \right) \,\,.
\end{equation}
For the next reduction we differentiate both sides of the identity $R_{g}^{*}\Omega^{4}=\hat{\Omega^{4}}$. We have
\begin{eqnarray}
R_{g}^{*}d\Omega^{4}&=&\hat{\alpha}_{44}\wedge \hO^{4}+\hT_{13}^{4}\hO^{1} \wedge \hO^{3}+\hT_{23}^{4}\hO^{2} \wedge \hO^{3}\hspace{.1in} (\mbox{mod}\hspace{.05in} \Omega^{4})  \nonumber \\
&=&\ba_{33}((A_{11}\hT_{13}^{4}+A_{12}\hT_{23}^{4})\Omega^{1}\wedge\Omega^{3}+(A_{21}\hT_{13}^{4}+A_{22}\hT_{23}^{4})\Omega^{2}\wedge\Omega^{3})\hspace{.1in} (\mbox{mod}\hspace{.05in} \Omega^{4}) \nonumber
\end{eqnarray} \,\,.
On the other hand
$$
d\hat{\Omega}^{4} = \ba_{44} d\Omega^{4}  
                     = \ba_{44}(T_{13}^{4}\Omega^{1}\wedge                \Omega^{3}+T_{23}^{4}\Omega^{2}\wedge \Omega^{3})\hspace{.1in} (\mbox{mod}\hspace{.05in} \Omega^{4}) \,\,\,.   
$$
The induced action of $G_{1}$ on the torsion plane $(T_{13}^{4},T_{23}^{4})$ is therefore
\begin{equation}
\left( \begin{array}{c} 
\hT^{4}_{13} \\ 
\hT^{4}_{23}
\end{array} \right) =
\frac{\epsilon}{a_{44}} A^{-1}
\left( \begin{array}{c} 
T^{4}_{13} \\ 
T^{4}_{23}
\end{array} \right) \,\,\,.  \label{eq:ACTION1}
\end{equation}
The torsion plane $(T_{13}^{4},T_{23}^{4})\neq (0,0)$ since $I^{(2)}=0$ implies that $d\Omega^{4}\wedge \Omega^{4}\neq 0$ and we have already established that $T_{12}^{4}=0$. We can therefore use the action to force $(T_{13}^{4},T_{23}^{4})= (0,1)$. The torsion $T_{23}^{4}\Omega^{2}\wedge \Omega^{3}$ can be interpreted as the (normalized) dual curvature of the rank three distribution $\D^{1}$. The statement that $(T_{13}^{4},T_{23}^{4})\neq (0,0)$ is equivalent to $\D$ not being integrable. To determine the subgroup that stabilizes this choice of torsion, we investigate 
\begin{equation}
R_{g}^{*}\left( \begin{array}{c} 
0 \\ 
1
\end{array} \right)
=
\frac{\epsilon}{a_{44}} A^{-1}\left( \begin{array}{c} 
0 \\ 
1
\end{array} \right)
=\left( \begin{array}{c} 
0 \\ 
1
\end{array} \right).
\end{equation}
As $A\in O(2)$ it must be of the form
\begin{equation}
\left( \begin{array}{cc} 
\epsilon_{1} \epsilon_{2} & 0 \\ 
0 & \epsilon_{2}
\end{array} \right)
\end{equation}
where $\epsilon_{1}, \epsilon_{2} \in \{-1,1\}$. We must also have $a_{44}=\epsilon_{1}\epsilon_{2}$ so that the stabilizer subgroup $G_{2}$ consists of matrices of the form
\begin{equation}
\left(
\begin{array}{cccc} 
\epsilon_{1}\epsilon_{2} &0 & 0 &B_{14}\\ 
0 & \epsilon_{2} & 0 & B_{24}\\
  0& 0  & \epsilon_{1} & a_{34}\\
  0& 0  & 0 & \epsilon_{1}\epsilon_{2}
\end{array} \right)  \label{eq:G1}
\end{equation}
where $\epsilon_{1}, \epsilon_{2} \in \{-1,1\}$ and $B_{14}$, $B_{24}$ and $a_{34}\in \Re$. We compute
\begin{equation}
R_{g}^{*} {\bf \Omega}
=
\left( \begin{array}{l} 
\hO^{1} \\ 
\hO^{2}\\
\hO^{3}\\
\hO^{4}
\end{array} \right)=
\left( \begin{array}{c} 
\epsilon_{1}\epsilon_{2}\Omega^{1}-B_{14}\Omega^{4}\\ 
\epsilon_{2} \Omega^{2}-B_{24}\Omega^{4}  \\
\epsilon_{1}\Omega^{3}-\epsilon_{2}a_{34}\Omega^{4}\\
\epsilon_{1}\epsilon_{2}\Omega^{4}
\end{array} \right) \,\,\,.
\end{equation}
The structure equations are now
\begin{eqnarray}
d\left( \begin{array}{l} 
\Omega_{1} \\ 
\Omega_{2}\\
\Omega_{3}\\
\Omega_{4}
\end{array} \right)
=
-\left(
\begin{array}{c} 
\beta_{14}\wedge \Omega^{4}\\ 
 \beta_{24}\wedge\Omega^{4}\\
 \alpha_{34}\wedge\Omega^{4}\\
0
\end{array} \right)
+
\left( \begin{array}{c}
T^{1}_{12}\Omega^{1} \wedge \Omega^{2} +
T^{1}_{13}\Omega^{1} \wedge \Omega^{3}+
T^{1}_{23}\Omega^{2} \wedge \Omega^{3} \\ 
T^{2}_{12}\Omega^{1} \wedge \Omega^{2} +
T^{2}_{13}\Omega^{1} \wedge \Omega^{3}+
T^{2}_{23}\Omega^{2} \wedge \Omega^{3}\\
\Omega^{1} \wedge \Omega^{2} +
T^{3}_{13}\Omega^{1} \wedge \Omega^{3}+
T^{3}_{23}\Omega^{2} \wedge \Omega^{3}\\
T^{4}_{14}\Omega^{1} \wedge \Omega^{4} +\Omega^{2} \wedge \Omega^{3}+T_{24}^{4}\Omega^{2}\wedge\Omega^{4}+T_{34}^{4}\Omega^{3}\wedge\Omega^{4}
\end{array} \right)  \,\,\,.
\end{eqnarray}
$B_{2}$ is not an $e$-structure so again we differentiate both sides of the identity $R_{g}^{*}\Omega=\hat{\Omega}$ to determine the action of $G_{2}$ on the torsion.
After some computation, we find that 
\begin{eqnarray}
d\hO^{1}&=&\epsilon_{1}\epsilon_{2}(T_{13}^{1}\Omega^{1}\wedge \Omega^{3}+T_{23}^{1}\Omega^{2}\wedge \Omega^{3}+T_{12}^{1}\Omega^{1}\wedge \Omega^{2})-B_{14}\Omega^{2}\wedge \Omega^{3} \hspace{.15in} (\mbox{mod} \hspace{.1in} \Omega^{4}) \nonumber \\
d\hO^{2}&=&\epsilon_{2}(T_{13}^{2}\Omega^{1}\wedge \Omega^{3}+T_{23}^{2}\Omega^{2}\wedge \Omega^{3}+T_{12}^{2}\Omega^{1}\wedge \Omega^{2})-\epsilon_{1} B_{24}\Omega^{2}\wedge \Omega^{3} \hspace{.15in} (\mbox{mod} \hspace{.1in} \Omega^{4}) \nonumber \\
d\hO^{3}&=&\epsilon_{1}(T_{13}^{3}\Omega^{1}\wedge \Omega^{3}+T_{23}^{3}\Omega^{2}\wedge \Omega^{3}+\Omega^{1}\wedge \Omega^{2})-\epsilon_{2} a_{34}\Omega^{2}\wedge \Omega^{3} \hspace{.15in} (\mbox{mod} \hspace{.1in} \Omega^{4}). \nonumber
\end{eqnarray}
Also,
\begin{eqnarray}
R_{g}^{*}(d\Omega^{1})&=&\epsilon_{2}\hat{T}_{13}^{1} \Omega^{1}\Omega^{3}+\epsilon_{1}\epsilon_{2}\hat{T}_{23}^{1}+\epsilon_{1}
\hat{T}_{12}^{1}\Omega^{1}\wedge \Omega^{2} \hspace{.15in} (\mbox{mod} \hspace{.1in} \Omega^{4})  \nonumber \\
R_{g}^{*}(d\Omega^{2})&=&\epsilon_{2}\hat{T}_{13}^{2} \Omega^{1}\Omega^{3}+\epsilon_{1}\epsilon_{2}\hat{T}_{23}^{2}+\epsilon_{1}
\hat{T}_{12}^{2}\Omega^{1}\wedge \Omega^{2} \hspace{.15in} (\mbox{mod} \hspace{.1in} \Omega^{4}) \nonumber \\
R_{g}^{*}(d\Omega^{3})&=&\epsilon_{1}\hat{T}_{13}^{3} \Omega^{1}\Omega^{3}+\epsilon_{1}\epsilon_{2}\hat{T}_{23}^{3}+\epsilon_{1}
\Omega^{1}\wedge \Omega^{2} \hspace{.15in} (\mbox{mod} \hspace{.1in} \Omega^{4}). \nonumber
\end{eqnarray}
Matching the $\Omega^{2}\wedge \Omega^{3}$ terms we find that 
\begin{eqnarray}
\hat{T}_{23}^{1}&=&T_{23}^{1}-\epsilon_{1}\epsilon_{2}B_{14}, \nonumber \\
\hat{T}_{23}^{2}&=&\epsilon_{1}T_{23}^{2}-\epsilon_{2}B_{24},   \\
\hat{T}_{23}^{3}&=&\epsilon_{2}T_{23}^{3}-\epsilon_{1}a_{34} \,\,\,. \nonumber
\end{eqnarray}
We can therefore use the action of $G_{1}$ to force $T_{23}^{1}=T_{23}^{2}=T_{23}^{3}=0$.
 The stabilizer subgroup $G_{final}$ for this choice of torsion consists of matrices of the form 
\begin{equation}
\left(
\begin{array}{cccc} 
\epsilon_{1}\epsilon_{2} &0 & 0 &0\\ 
0 & \epsilon_{2} & 0 & 0\\
  0& 0  & \epsilon_{1} & 0\\
  0& 0  & 0 & \epsilon_{1}\epsilon_{2}
\end{array} \right)  \,\,\,.
\end{equation}
{\it The reduced structure group is discrete} so we now have an $e$-structure $B_{final}$. The tautological one-form $(\Omega_{1},\Omega_{2},\Omega_{3},\Omega_{4})^{\mbox{tr}}$  provides a full  coframing for   $B_{final}$. The $B_{final}$ structure equations are
\begin{equation}
d\left( \begin{array}{l} 
\Omega_{1} \\ 
\Omega_{2}\\
\Omega_{3}\\
\Omega_{4}
\end{array} \right)
=
\left(
\begin{array}{cccccc} 
T_{12}^{1} &T_{13}^{1} & T_{14}^{1} &0& T_{24}^{1}&T_{34}^{1}\\ 
T_{12}^{2} &T_{13}^{2} & T_{14}^{2} &0& T_{24}^{2}&T_{34}^{2}\\
1& T_{13}^{3}  & T^{3}_{14} & 0&T_{24}^{3} &T_{34}^{3}\\
  0& 0  & T_{14}^{4} & 1 & T_{24}^{4} & T_{34}^{4}
\end{array} \right)
\left(
\begin{array}{c} 
\Omega^{1} \wedge \Omega^{2} \\ 
\Omega^{1} \wedge \Omega^{3} \\
\Omega^{1} \wedge \Omega^{4} \\
\Omega^{2} \wedge \Omega^{3} \\
\Omega^{2} \wedge \Omega^{4} \\
\Omega^{3} \wedge \Omega^{4} 
\end{array} \right)
\end{equation}
 where the $T_{ij}^{k}$ are functions on $B_{final}$.
What remains is to determine any second order relations between the torsion functions. To determine these we use the fact that $d^{2}=0$. After some computation, we find that $T_{14}^{4}=T_{12}^{2}+T_{13}^{3}$. 
We summarize these results in the following theorem:

\begin{theorem} \label{structure}
Associated to any nonholonomic Engel structure $\{Q, \g=\langle \hspace{.05in} \cdot \hspace{.05in} , \hspace{.05in} \cdot \hspace{.05in}\rangle , \D \}$ there is a canonical $G\cong {\bf Z}_{2}\times {\bf Z}_{2}$-structure $B_{final}$. The tautological one-form $(\Omega_{1},\Omega_{2},\Omega_{3},\Omega_{4})^{\mbox{tr}}$ provides a canonical coframing for $B_{final}$. The $B_{final}$ structure equations are
\begin{equation}
d\left( \begin{array}{l} 
\Omega_{1} \\ 
\Omega_{2}\\
\Omega_{3}\\
\Omega_{4}
\end{array} \right)
=
\left(
\begin{array}{cccccc} 
T_{12}^{1} &T_{13}^{1} & T_{14}^{1} &0& T_{24}^{1}&T_{34}^{1}\\ 
T_{12}^{2} &T_{13}^{2} & T_{14}^{2} &0& T_{24}^{2}&T_{34}^{2}\\
1& T_{13}^{3}  & T^{3}_{14} & 0&T_{24}^{3} &T_{34}^{3}\\
  0& 0  & T_{12}^{2}+T_{13}^{3} & 1 & T_{24}^{4} & T_{34}^{4}
\end{array} \right)
\left(
\begin{array}{c} 
\Omega^{1} \wedge \Omega^{2} \\ 
\Omega^{1} \wedge \Omega^{3} \\
\Omega^{1} \wedge \Omega^{4} \\
\Omega^{2} \wedge \Omega^{3} \\
\Omega^{2} \wedge \Omega^{4} \\
\Omega^{3} \wedge \Omega^{4} 
\end{array} \right)
\end{equation}
\end{theorem}
 According to the {\it framing lemma} (lemma \ref{framing}) the largest Lie group of symmetries of a nonholonomic structure on an Engel manifold is the dimension of $B_{final}$ which is four. In this case the $T^{I}_{JK}$ are constants and can be identified with the structure constants of the four-dimensional Lie algebra of the symmetry group. The Jacobi identities are obtained using the identity $d^{2}=0$. We have computed these, and it appears that the set of possible symmetry algebras form a rather complicated subvariety of the variety of all four-dimensional Lie algebras. We leave as an open problem the classification of all possible four-dimensional symmetry algebras for nonholonomic structures on an Engel manifold.

\paragraph{The rolling penny (Continued).} An example of a structure with maximal symmetry is given by the rolling penny. A $B_{final}$-adapted coframe for the penny-table system is 
\begin{eqnarray}
\eta^{1}&=&\sqrt{\frac{ma^{2}+I}{2}}d\phi \nonumber \\
\eta^{2}&=&\sqrt{\frac{J}{2}}d\theta   \\
\eta^{3}&=&\frac{\sqrt{J(ma^{2}+I)}}{2}(-\sin \theta dx+\cos \theta dy) \nonumber \\
\eta^{4}&=&\sqrt{\frac{m}{2}}(\cos \theta dx+\sin \theta dy-d\phi)
\nonumber
\end{eqnarray}
The structure equations are
\begin{eqnarray}
d\eta^{1}&=&0 \nonumber \\
d\eta^{2}&=&0   \\
d\eta^{3}&=&\eta^{1}\wedge \eta^{2}-\sqrt{\frac{ma^{2}+I}{m}} \hspace{.1in} \eta^{2}\wedge \eta^{4} \nonumber \\
d\eta^{4}&=&\frac{2}{J} \hspace{.1in}  \sqrt{\frac{m}{ma^{2}+I}} \hspace{.1in} \eta^{2}\wedge \eta^{3} \,\,. \nonumber 
\end{eqnarray}
The torsion functions are constant so by the framing lemma (lemma \ref{framing}) we can identify these constants with the structure constants Lie group of symmetries of this system. We recognize them as the structure constants for the Lie algebra of the group $SE(2)\times SO(2)$ which is isomorphic to the configuration space of the penny-table system.  
 
\paragraph{$B_{final}$-adapted frames and coframes.} The $e$-structure $B_{final}$ has a canonical coframing which descends to a coframing and hence a framing, up to signs, on $Q$. There should be a relationship between this framing and a canonical line field possessed by any Engel manifold. In this section we briefly describe this relationship. If $Q$ and $\D$ are both oriented, then $Q$ is parallelizable  and the following constructions can be made globally (\cite{Montgomery}).

Let  $\eta$ be a  $B_{final}$ adapted coframe on $U\subset Q$ with dual frame $X=\{X_{I}\}$ defined by $\eta^{I}(X_{J})=\delta_{IJ}$. If $\bar{\eta}$ is any other $B_{final}$-adapted coframe with dual frame $\bar{X}$on $U$, then, by theorem \ref{structure}, $\bar{\eta}$ is related to $\eta$ by $\bar{\eta}^{1}=\epsilon_{1} \epsilon_{2}\eta^{1}$,  $\bar{\eta}^{2}= \epsilon_{2}\eta^{1}$, $\bar{\eta}^{3}= \epsilon_{1} \eta^{3}$,  $\bar{\eta}^{4}=\epsilon_{1} \epsilon_{2}\eta^{4}$. The dual frames are related in precisely the same way: $\bar{X}_{1}=\epsilon_{1}\epsilon_{2}X_{1}$, $\bar{X}_{2}=\epsilon_{2}X_{2}$,  $\bar{X}_{3}=[\bar{X}_{1}, \bar{X}_{2}]=\epsilon_{1}X_{3}$, and $\bar{X}_{4}=[\bar{X}_{2},\bar{X}_{3}]=\epsilon_{1}\epsilon_{2}X_{4}$.

An important feature of an Engel distribution is the presence of  a canonical line field $L\subset \D$ (\cite{Montgomery}, \cite{Montgomery1}). $L$ is defined by the condition that $[L,\D^{1}]\subset \D^{1}$. Here we are abusing notation, using $L$ for the line field or a vectorfield spanning $L$. We have
\begin{corollary}
Let $\eta=\eta^{I}$ be a $B_{final}$-adapted coframe. Let $X=\{X_{I} \}$ be the dual frame defined by $\eta^{I}(X_{J})=\delta_{IJ}$, then $L=\mbox{span}(X_{1})$.
\end{corollary}
\noindent {\bf Proof.} Suppose $L$ is spanned by the vectorfield $Y=aX_{1}+bX_{2}$. Since $\eta^{4}$ annihilates $\D^{1}$ we have $\eta^{4}([X_{3},Y])=0$. Then$$0=\eta^{4}([X_{3},Y])=X_{3}\eta^{4}(Y)-Y\eta^{4}(X_{3})-d\eta^{4}(X_{3},Y)=-d\eta^{4}(X_{3},Y).$$ But $d\eta^{4} \equiv \eta^{2} \wedge \eta^{3} \hspace{.1in} \mbox{mod}\hspace{.05in} (\eta^{4})$ so we must have
\begin{eqnarray} 
0=\eta^{2}\wedge \eta^{3}(X_{3},Y)&=& \eta^{2}(X_{3})\eta^{3}(Y)-\eta^{3}(X_{3})\eta_{2}(Y) \nonumber \\
&=&-\eta^{3}(X_{3})\eta_{2}(Y) \nonumber \\
&=&-b. \nonumber
\end{eqnarray}
$L$ is therefore spanned by $X_{1}$.
This concludes the arguement.

There is a natural metric, associated with $B_{final}$, on $Q$ given by $\ g_{nat}=\tilde{\eta}^{1}\otimes \tilde{\eta}^{1}+\cdot \cdot \cdot  \tilde{\eta}^{4}\otimes \tilde{\eta}^{4}$ where $\tilde{\eta}$ is any $B_{final}$-adapted coframe.  Clearly all $B_{final}$-adapted coframes induce this same metric; using the subriemannian metric $\ g_{nat}|_{\D}$ we  form $L^{\perp}$ within $\D$ so that $\D =L\oplus L^{\perp}$. By construction, $X_{2}$ spans $L^{\perp}$. 

  
\section{Nonholonomic dynamics: Chaplygin Hamiltonization} \label{GChap}

Historically, Hamiltonization 
of nonholonomic systems 
started with Chaplygin's {\it last multiplier method}. 
In the new time, the dynamics obeys Euler-Lagrange equations
{\it without
extra terms}; the gyroscopic force (\ref{NHforce}) ``magically'' disappers!
When  after a time reparametrization
the compressed system can be
described as a Hamiltonian system,  symplectic techniques can be employed. 
A number of NH systems have been Hamiltonized, and   some interesting ones are
Liouville-integrable, see    \cite{Veselovs},  \cite{Kozlov}, \cite{Fedorov1},
 \cite{Fedorov2}, \cite{Dragovic}, \cite{Borisov}, \cite{Borisov1}, \cite{Borisov2},
 \cite{Borisovsphere}, \cite{Jovanovic}.

\subsection{Compression to $T^*S, \,S = Q/G$; existence of invariant measures}

  We recall from the introduction that the compressed   system  has a concise almost Hamiltonian
  $$dH^{\phi} = i_{X_{NH}} \,\Omega_{NH} \,\,\,, \,\,\, \Omega_{NH} := \Omega_{can}^{T^*S} + {\rm (J.K)} \,\,\,\,,\,\,\,\,  d\Omega_{NH} \neq 0 \,\,({\rm in}\,{\rm general})\,,
$$
where $\Omega_{can}^{T^*S}$ is the canonical 2-form of $T^*S$ and the ${\rm (J.K)}$ term is
a semi-basic two form, which in general is non-closed. It combines  the momentum  $J$ of the $G$-action on $T^*Q$, and the curvature $K$ of the connection. As this is important for the
remaining, we outline the derivation (see \cite{Koillerromp} for details).  Given the coframe coordinates $m,\,,\,\epsilon(q)$ in $T^*Q$ (see \ref{mepsilon}) the Poisson
bracket matrix relative to 
$\, \epsilon_I \,, \,\,  dm_I \,\,$ is
\begin{equation} \label{poisson}  [\Lambda] = [\Omega]^{-1} =
\left(
\begin{array}{ll}   0_n & I_n\\   -I_n & \,\,\,\,E
\end{array} \right)
\end{equation}
with
\begin{equation} \label{R}
 E_{JK} = m_I d\epsilon_I
(e_J,e_K) = - m_I
\epsilon_I[e_J,e_K]  \,\,.
\end{equation}
Let us consider the case of a principal bundle  $\pi: Q^n \rightarrow S^s$  with Lie group $G^r$ acting on the left,   $ r = n-s$.
Recall our convention: capital roman letters
$I,J,K,$ etc., run from $1$ to $n$.
 Lower case roman characters $i,j,k $ run from $1$ to $s$.  Greek
characters $\alpha, \beta, \gamma$, etc., run from  $s+1$ to $n$.

Fix a connection $\lambda = \lambda(q): T_q Q \rightarrow Lie(G)$ defining a
$G$-invariant distribution $\mathcal{H}$ of horizontal subspaces.
Denote by  $K(q) = d\lambda \circ {\rm Hor}: T_q Q \times T_q Q \rightarrow
\mathcal{G}$ the curvature 2-form (which is, as
well known,
$Ad$-equivariant).
Choose a local frame $\overline{e}_i$ on $S$.  
For simplicity, we may assume that
\begin{equation} \label{easychoice}
\overline{e}_i = \partial/\partial
s_i
\end{equation} are the coordinate vectorfields of a
chart
$s: S
\rightarrow \Re^s$.

Let  $e_i = h(\overline{e}_i)$ their horizontal
lift to $Q$.  We complete to a moving frame
of $Q$ with vertical vectors $e_{\alpha}$
which we will specify in a moment.
The dual basis will be denoted $\epsilon_i, 
\epsilon_{\alpha}$ and we write $ p_q = m_i \epsilon_i + m_{\alpha}
\epsilon_{\alpha}$. These are in a sense the ``lesser moving'' among all the moving frames
adapted to this structure.
We now describe how the $n \times n$ matrix $ E =
(E_{IJ})$ looks like in this setting.\\

i) The  $s \times s$ block  $(E_{ij})$.\\

Decompose 
$[e_i,e_j] =
h[\overline{e}_i,\overline{e}_j] + V[e_i,e_j] =
V[e_i,e_j]
$ 
into
vertical and horizontal parts.  The choice
(\ref{easychoice}) is convenient, since $ \overline{e}_i$ and $\overline{e}_j$  commute:
$[e_i,e_j]$ is vertical. Hence
 \begin{equation} E_{ij} =  - p_q[e_i,e_j] = - m_{\alpha}
\epsilon_{\alpha} [e_i,e_j] \,\,.
\end{equation}

Now by Cartan's rule,  
$$ K(e_i,e_j) = e_i \lambda(e_j) - e_j
\lambda(e_i) -
\lambda [e_i,e_j] =  -
\lambda [e_i,e_j]
\in
\mathcal{G}
$$
Thus we showed that  
\begin{equation} [e_i,e_j]_q = - K(e_i,e_j) \cdot q
\end{equation}
Moreover, let
$ J: T^*Q \rightarrow Lie(G)^* $ the momentum mapping. We have
$$ (J(p_q), K_q(e_i,e_j) ) = p_q \,(
K(e_i,e_j).q\,) = - p_q [e_i,e_j] \,\,\,\,\,\,(= E_{ij})
$$
\begin{theorem}  (The  J.K formula)
\begin{equation}  E_{ij} = (J(p_q), K_q(e_i,e_j) )  
\end{equation}
\end{theorem}
This gives a nice   description for   this block, under the choice
$[\overline{e}_i,\overline{e}_j] = 0.$
Notice that the functions
$E_{ij}$ depend on $s$ and the components
$m_{\alpha}$, but do not depend on $g$. This is
because the $Ad^*$-ambiguity of the momentum
mapping $J$ is cancelled by the $Ad$-ambiguity
of the curvature $K$.  The other blocks are not needed here, but we include for completeness.\\

ii) The  $ r \times r$ block  $(E_{\alpha \beta})$. \\

Choose a basis $X_{\alpha}$ for $ Lie(G)$. We take   $e_{\alpha}(q) = X_{\alpha} \cdot q$  as the  vertical distribution.  Choosing a point $q_o$ allows
identifying  the Lie group $G$ with the fiber containing $G q_o$, so that ${\rm id} \mapsto q_o$.
Through the mapping $g \in G \mapsto gq_o \in Gq_o$ the vectorfied $e_{\alpha}$ is identified
to a {\it right} (not left!) invariant vectorfield in $G$.  The commutation relations
for the $e_{\alpha}$  $  [e_{\alpha},e_{\beta}] = - c_{\alpha \beta}^{\gamma} \, e_{\gamma}$ 
appear with a minus sign. Therefore
\begin{equation}
E_{\alpha \beta} = m_{\gamma} c_{\alpha \beta}^{\gamma} \,\,. 
\end{equation}

iii) The  $s \times n$ block $(E_{i \alpha})$.  \\

The vectors  $[e_i, e_{\alpha}]$ are vertical, but their values depend on the
 specific  principal bundle one is working with, and there are some noncanonical choices.  Given a section $\sigma: U_S \rightarrow Q$ over the
coordinate chart  $s: U_S \rightarrow \Re^m$ on $S$, we need to know the coefficients
$b^{\gamma}_{i \alpha}$ in the expansion
$$ [e_i, e_{\alpha}](\sigma(s)) = b^{\gamma}_{i \alpha} (s) \, e_{\gamma} \,\,.$$
Then
\begin{equation}
 E_{i \alpha}(\sigma(s))  = - m_{\gamma}\, b^{\gamma}_{i \alpha} (s) \,\,.
\end{equation}
At another point on the fiber, we need the adjoint representation  $Ad_g: Lie(G)
\rightarrow Lie(G), \, X \mapsto g_*^{-1} X g, \,$  described by a matrix
$(A_{\mu \alpha}(g))$ such that
\begin{equation} Ad_g (X_{\alpha}) =  A_{\mu \alpha}(g) X_{\mu} \,\,\,.
\end{equation}
Then
\begin{equation}
[e_i, e_{\alpha}](g \cdot \sigma(s)) = - m_{\gamma} b^{\gamma}_{i \mu} (s) A_{\mu \alpha}(g) \,\,.
\end{equation}

\paragraph{The clockwise diagram.}  Starting on  $p_s \in T^*S$ we go clockwise to  $P_q \in Leg({\cal H}) \subset T^*Q$,
for some $q$  on the fiber $\pi^{-1}(s)$ of $Q$ over $s$. 
 \begin{equation} \begin{array}{lll}  \label{diagram} 
 {\cal H} \subset TQ & \longrightarrow & Leg({\cal H}) \subset T^*Q \\
& Leg &   \\
\, \uparrow & & \,\,  \\
\, h & &\,\,  \\
\,\,  | &         & \,\,   \\
& & \\
TS & \longleftarrow & T^*S \\
& (Leg^{\phi})^{-1} &
\end{array}
\end{equation}
Taking differentials of all maps in (\ref{diagram}) we obtain an induced principal connection
$\hat{\phi}$  
in  the bundle  $G \hookrightarrow Leg({\cal H}) \rightarrow T^*S \,.$
 Let  $v,w,z \in T_{p_s}(T^*S)$,
$\,\,\, V,W,Z$ horizontal lifts at $P_q \in Leg({\cal H})$, 
and denote by $\hat{K}$ the curvature of this induced connection.
The following  proposition is basically a rephrasing of a result in \cite{BatesSniatycki}.
\begin{proposition}
\begin{equation}
d\,{\rm (J.K)}(v,w,z) = {\rm cyclic} (dJ(V),K(W,Z)) \,\,\,.
\end{equation}
\end{proposition}

\paragraph{Densities of invariant measures and a dimension dependent exponent.}  
 A  necessary and sufficient
condition for the existence of an invariant measure for compressed Chaplygin systems was obtained by \cite{Cantrijn} 
(Theorem 7.5).  Since in $T^*S$ there is a natural Liouville measure $d{\rm vol} = ds_1 ... ds_m dp_1 ... dp_m$, where $(s,p)$ are 
 coordinates in $T^*S$, the
{\it   density function} $F$ {\it produces an educated guess}  {\it for a time reparametrization} which may Hamiltonize the compressed system. If ${\rm dim}(S) = m$ and $f\,\Omega_{NH}$ is closed,   the time-reparametrized
vectorfield $X_{NH}/f$ has
the invariant measure $f^m\,d{\rm vol}.$    $X_{NH}$ 
will have the invariant measure $f^{m-1}\, ds_1 ... ds_m dp_1 ... dp_m$. Working backwards, 
if a measure density $F$ is known so that
$ F(s) d{\rm vol}$ is an invariant measure for $X_{NH}$,  then the obvious {\it candidate} 
for conformal factor is
\begin{equation} \label{dimensioncount}
f = F(s)^{\frac{1}{m-1}} \,\,.
\end{equation}  
This dimension dependent exponent will be relevant in the
Chaplygin marble, see section \ref{Chaplyginspheresection}. 

\paragraph{Invariant measures for LR systems.}  Let $Q=G$ a unimodular Lie group and identify $TG \equiv T^*G$ via the bi-invariant metric.  Assume that $H \subset G$ is a subgroup acting on the left and preserving the distribution: $ {\cal D}_{hg} = h {\cal D}_g = h \,{\cal D}g $ (which boils down
to $Ad_{h^{-1}}\, {\cal D} = h^{-1} {\cal D} h = {\cal D}$). 
The   Legendre transform $Leg: Lie(G) \rightarrow Lie(G) \equiv Lie^*(G)$ of a natural, left invariant  Lagrangian,  is represented by a positive symmetric transformation $A: Lie(G) \rightarrow Lie(G)$,
the {\it inertia operator}. 

For each $g \in G$, let  $ P^1_g $ and $P^2_g $ be, respectively, the projections  of $Lie(G)$  relative to the
decomposition  $Lie(G) = Ad_{g^{-1}} Lie(H)   \oplus Ad_{g^{-1}}{\cal D}  $.
 We can also think of $P^2_g$ as a map  $P^2_g: T_g\,G \rightarrow  {\cal D}\, g$,   projection  parallel
to the vertical spaces $Lie(H)g$. Let  $  P^2_g \,o\, Leg_g : {\cal D}g \rightarrow {\cal D}g$.  This map descends to the compressed Legendre transform $Leg_s^{\phi}: T_s\,S \rightarrow
T_s \,S \equiv T_s^*S$, where $S =G/H$ is the homogeneous space whose metric is induced by the
bi-invariant metric on $G$.  Consider the function 
\begin{equation}
F(s) = {\rm det} Leg_s^{\phi} \,\,\,.
\end{equation}
The following result is a rephrasing of a theorem by \cite{Veselovs}, see also
\cite{Fedorov1} (Theorem 3.3)\footnote{We do not need to assume
${\cal D}$ and $Lie(H)$ to be orthogonal with respect to the bi-invariant metric.}.

\begin{theorem} \label{Veselovstheorem} The reduced LR-Chaplygin system  in the homogeneous space $T^*(G/H)$ always has the
invariant measure
\begin{equation}  \label{measureLR}
 \nu = F(s)^{-1/2} \, ds_1 \cdots ds_m \,dp_1 \cdots dp_m \,\,\,,\,\,\,\,\,\,  F(s) = {\rm det}\, Leg_s^{\phi} \,\,.
\end{equation}
The density can be also calculated by  the ``dual''  formula
\begin{equation} \label{dualmeasure}
 F(s) = {\rm det}(A) \, {\rm det}\left( P^2_g \,o\, A^{-1}|_{g^{-1}Lie(H)g} \right)
\end{equation}
($P^1_g$ is the projection over $g^{-1}Lie(H)g$ parallel to $g^{-1} {\cal D} g)$.
\end{theorem}
The second formula may be easier to use if there are few constraints.

 \paragraph{Almost  Hamiltonian systems.}  Let $\Omega$  be a non-degenerate   (but in general, non-closed) 2-form on  $M^{2n}$, and $H$ be
 a function   on $M$.  Denote (as usual) by $X=X_H$  the skew-gradient vectorfield
defined by $i_X \Omega = dH$. 
 We say $X_H$ is {\it almost Hamiltonian}. If $\alpha$ is a closed 1-form,
 the vectorfield $X = X_{\alpha}$ defined by $i_X \Omega = \alpha$ is called
 {\it locally almost  Hamiltonian}. Distilling a construction in  \cite{Stanchenko}, we  formalize an extension of  the notion of a conformally symplectic structure.

 The 2-form $\Omega$ is called H (or $\alpha$)- {\it affine symplectic} if there is a function
$f > 0 $ on $M$ and a two form $\Omega_o$ such that i) $ \,\,i_{X} \Omega_o \equiv 0  \,\,\,  $; ii)
$ \,\, \Omega - \Omega_o  $ is non-degenerate,
and iii) $\,\, \tilde{\Omega} = f (\Omega - \Omega_o ) $ is closed\footnote{We must 
admit, however,  that we found no example yet where the affine term is really needed.
This notwithstanding, at any point where $X \neq 0$,  the contraction condition yields $d = 2n$ equations on $d(d-1)/2$ unknowns (local coordinate coefficients of $\Omega_o$). This allows  additional freedom   to Hamiltonize $X$ rather than just requiring conformality of $\Omega$.}.

The first condition implies that $X$ does not ``see'' $\Omega_o$. Together with the third, we get  $\tilde{\Omega}(X/f,\bullet) = dH$ so  the vectorfield $X/f$ is (truly) Hamiltonian with respect
to the symplectic form  $ \tilde{\Omega} $. 



The closedness condition can be restated as 
\begin{equation} \label{affinely}
d (\Omega - \Omega_o ) = (\Omega - \Omega_o ) \wedge \theta\,\,,\,\,\, 
{\rm where }\,\,\, \theta = df/f \,\,\,.
\end{equation}
When (\ref{affinely}) holds with $\alpha$ a closed 
(but not necessarily exact) 1-form, we say that $ \Omega  $ is {\it locally} affine  symplectic.
The following proposition describes the obstruction to Hamiltonization once $f$ is given.

\begin{theorem} \label{criterionHamiltonizable} Given a locally almost hamiltonian system ($\Omega,\alpha$) and an educated guess $f>0$, an
affine term $\Omega_o$ exists with $d(f\Omega - \Omega_o) = 0$  if and only if $\,i_X \,d(f \Omega) = 0 $.  
\end{theorem}
The proof is quite easy. The vectorfield $X$ satisfies $ i_X \Omega = \alpha$. Since the same equation holds by replacing $X$ by $X/f$ and $\Omega$ by $f \Omega$,  to expedite notation we may assume
$f \equiv 1$.  Let us prove that $\Omega_o$ exists if $ i_X d\Omega = 0$.  Since $d(i_X \,\Omega) = d \alpha = 0$, we see that the Lie derivative  $L_X \Omega = 0$.
Consider a regular point of $X$. By the flow box theorem there are coordinates so that
$ X = \partial/\partial x_1$. Since $L_X \Omega = 0$, the coefficients of this 2-form do not
depend on the coordinate $x_1$ (but there may exist terms with a $dx_1$ factor).  However,
our hypothesis $i_{\partial/\partial x_1} \, d\Omega = 0$ ensures that there are no terms
containing a $dx_1$ factor in $d\Omega$.  Thus $d\Omega$ can be thought as a 3-form in
the space of the remaining coordinates. By Poincare's theorem  $d\Omega = d\Omega_o$,
where $\Omega_o$ is a 2-form in the space of the remaining coordinates.  Hence
$i_X \Omega_o = 0$ and $d(\Omega - \Omega_o) = 0$, as desired. The converse is even easier.


\subsection{Examples: Veselova's system and Chaplygin spheres (marble or rubber)} \label{Chaplyginspheresection}

\cite{Veselovs1, Veselovs} considered one of the simplest   nonholonomic LR-Chaplygin system,   
$Q = SO(3)$ with a left invariant metric 
$L = T = \frac{1}{2} (A \Omega, \Omega)$, and subjected to a right invariant constraint which,  without loss of generality, can be assumed to be $\rho_3 = 0$.  Hence the
admissible motions satisfy $\omega_3 = 0$, where $\omega$ is the angular velocity
viewed in the space frame. This is a LR Chaplygin system on $S^1 \hookrightarrow SO(3) \rightarrow S^2$. 

{\it Chaplygin's  ball}  is a  sphere of radius $r$ and mass $\mu$, whose  center of mass is assumed to be at the geometric center, but the inertia matrix   $A = {\rm diag} (I_1,I_2,I_3)$ may have unequal entries. Thus its Lagrangian is given by
$ 2L = (A\Omega,\Omega) + \mu (\dot{x}^2 + \dot{y}^2 + \dot{z}^2)\,$. The configuration space  is the euclidian group $Q = SE(3)$.

In the case of the {\it marble}, the ball   rolls without slipping  on a horizontal 
plane, with rotations about the z-axis allowed\footnote{\cite{Chaplygin1} showed that the 3d problem
is integrable using elliptic coordinates in the sphere;   for $n > 3$ the problem is open. For basic informations, see \cite{Fedorov}, p. 147-149,
on the 3-d case and p. 153-156 for the general $n$-dimensional case.   For a detailed account on the algebraic integrability
of ``Chaplygin's Chaplygin sphere'', see \cite{Duistermaat}. \cite{Schneider} analyzed  control theoretical
aspects.}.  Thus the  distribution of admissible velocities  is defined by 
${\cal D}:\,\,\,\dot{z} = 0\,\,,\,\, \dot{x} = r \omega_1\,,\,\dot{y} = - r \omega_2\,\,.$
Both Lagrangian and constraints are preserved under the action of the euclidian
 motions in the plane, together with the vertical translations. $G = SE(2) \times \Re $ acts
on $Q$ via  
   $$(\phi,u,v,w).(R,x,y,z) = (S(\phi)R, e^{i\phi}(u+iv), z+w) \,\,.$$
The dynamics could be be  directly reduced to ${\cal D}/G$, see e.g. \cite{Blochnonlinearity},
but we will proceed in two stages. First, we Chaplygin-compress the dynamics from $TQ$ to $TSO(3)$ using the translation subgroup of $SE(3)$,    regarding  the 
constraint distribution as an abelian {\it connection} on
$Q$ with base space   $S = SO(3)$ and fiber  $ \Re^3$; the connection form is given by
\begin{equation} \label{connectionmarble}
  \alpha_{marble} := (dx - r \rho_2\,, \,dy + r \rho_1, \,dz)\,\,.
\end{equation}
There is {\it another}  $S^1$  action on $Q$, this time acting on the first factor only:
$  e^{i \phi}(R,z) = (S(\phi)R,z)$.  This action preserves the Lagrangian but does
{\it not} preserve the distribution:  $\,\,D_{ (S(\phi)R,z)} \neq 
 e^{i \phi}_* \,D_{(R,z)}$.  However,   its infinitesimal action is given by the right vectorfield $X_3^{r} \in {\cal D}$. Noether's theorem applies, so  $p_{\phi} = \ell_3$ is a constant of motion.  Therefore Chaplygin's marble equations  can  be reduced,
 on each level set $\ell_3$, to $T(SO(3)/S^1) = TS^2$. 

 In the case of {\it Chaplygin's rubber ball}\footnote{This problem was not studied by Chaplygin. For the
 physical justification, see \cite{NF} and \cite{Manoloprep}. As far as we know its
 integrability has not yet been established. Formally, Veselova's system is the limit of Chaplygin's rubber ball as $ r \rightarrow 0$.},   rotations about the vertical axis are forbidden (since such rotations would cause
 energy dissipation). Here the constraints are defined by
a sub-distribution  ${\cal H} \subset {\cal D}$  with  
  Cartan's 2-3-5 growth numbers and in fact
    defining  a connection
  on  $ SE(2) \times \Re \hookrightarrow Q \rightarrow S^2$ with 1-form  
  \begin{equation} \label{connectionrubber}
  \alpha_{rubber} := (\rho_3 \, \hat{k}\,,\, dx - r \rho_2\,, \,dy + r \rho_1\,,dz)\,\,.
\end{equation}
 
 \paragraph{The extrinsic viewpoint} For clarity we present the classical, direct derivation of the equations of motion, following the ``extrinsic viewpoint'' advocated by the Russian Geometric Mechanics school (\cite{Borisov}).
\smallskip

$\bullet$  For the rubber Chaplygin ball (and Veselova's):
  in the space frame one has $\dot{\ell} = \tau$, where $\tau = \lambda \hat{k}$ is the torque exerted by the constraint force.    The torque is
 vertical   because
  $(\tau, \omega) = 0$
for all $\omega$ with third component equal to zero.  Viewed in the body frame, 
\begin{equation} \label{Veselova1}
\dot{L} + \Omega \times L = \lambda \gamma \,\,\,,\,\,\, 
\end{equation}
 Together (\ref{1/2}), one gets a closed system of ODEs in
the space $(L,\gamma) \in \Re^3 \times \Re^3$, {\it provided} the relation between $\Omega$
and $L$ is obtained. In Veselova's example, $\Omega = A^{-1} L$. The multiplier can be eliminated by differentiating
the constraint equation $(\Omega,\gamma) = 0$.  
After a simple computation, one gets
\begin{equation} \lambda = \frac{(L, A^{-1}\gamma \times A^{-1} L)}{(\gamma,A^{-1}\gamma)}\,\,.
\end{equation}
Besides the standard integrals of motion  $2H = (A^{-1}L,L), \,\,(\gamma,\gamma)=1, \,\,
(A^{-1}L,\gamma) = 0$,  \cite{Veselovs} showed that there is a quartic polynomial integral
\begin{equation} G = (L,L) - (L,\gamma)^2
\end{equation}
and an invariant measure\footnote{The level sets of the four integrals are 2-tori, since there are no fixed points
in the dynamics. The existence of an invariant measure in the tori allows the explicit integration
via Jacobi's theorem. \cite{Veselovs} found a
´´rather unexpected connection with Neumann's problem''.}
\begin{equation}
\mu = f(\gamma) dL_1 \wedge dL_2 \wedge dL_3 \wedge d\gamma_1 \wedge d\gamma_2 \wedge d\gamma_3\,\,\,,\,\,\,  f(\gamma) =  (A^{-1}\gamma,\gamma)^{-1/2}\,\,\,.
\end{equation}

\smallskip

$\bullet$ For Chaplygin's marble: the angular momentum {\it at the contact point}, in the space frame $\vec{\ell}$   is constant. An engineer would argue that   both gravity and friction
produce no torque at that point; a mathematician would use the fact that the admissible vectorfields $V_i \in \mathcal{H}$ given by 
\begin{equation} 
V_1 := - r \, \partial/\partial y + X_1^{{\rm right}}\,\,,\,\, V_2:= r \,\partial/\partial x +  X_2^{{\rm right}}\,\,, \,\, V_3 := X_3^{{\rm right}} 
\end{equation} 
preserve the Lagrangian, and  would  invoke   NH-Noether's theorem.  Whichever explanation chosen, differentiating  $RL = \ell = RL$ and  $R \gamma = k$ one   gets Chaplygin's equations
\begin{equation}  \label{Chaplyginequations}
\dot{L} =   - \Omega \times L \,\,,\,\,\,\dot{\gamma} = - \Omega \times \gamma \,\,.
\end{equation}
  These two
 form a coupled system, since again $\Omega$ {\it is a linear function of} $L$ {\it depending
only on} $\gamma$:
\begin{equation} \label{LOmega}
  L = L_{\gamma}\,(\Omega) = A \Omega + \mu r^2 \gamma \times (\Omega \times \gamma) = \tilde{A} \Omega - \mu r^2 (\gamma,\Omega) \gamma\, \,\,,\,\,\,  \tilde{A} := A + \mu r^2 {\rm id}\,\,
\end{equation}  
A simple way to get this map is  to look at the total energy
\begin{equation}
 2T  =   (\omega, \ell) =  (\Omega, L) =   (A \Omega, \Omega)
+ \mu   (\dot{x}^2 + \dot{y}^2)  =   (A \Omega, \Omega)
+ \mu r^2 (\omega_1 ^2 + \omega_2^2) 
\end{equation}
which can be also written as
\begin{equation}  \label{compressedenergy}
2T = (\Omega, L) = (A \Omega, \Omega)
+ \mu r^2  ( \Omega \,,\, \gamma \times ( \Omega \times \gamma)\,) =
(\, \Omega\,,\, A \Omega + \mu r^2 \gamma \times (\Omega \times \gamma)\,)\,) \,\,.
\end{equation}
The expression $\gamma \times (\bullet \times \gamma)$ represents the projection
in the plane perpendicular to $\gamma$, and we get (\ref{LOmega}).
An {\it ansatz} for the inverse of the map (\ref{LOmega}) is (\cite{Duistermaat}),
\begin{equation} \label{inverse}
 \Omega = \Omega(L,\gamma) =  (L_{\gamma})^{-1}(L) =  \tilde{A}^{-1}L + \alpha(L) \tilde{A}^{-1}(\gamma) \,\, 
\end{equation}
and  one gets the interesting expression for $\alpha(L)$ (which will be used in equation (\ref{lomega}) and Proposition \ref{marbleconformal}):
\begin{equation}  \label{OmegaL}
 \alpha(L) = \mu r^2 \, \frac{(\gamma, \tilde{A}^{-1} L)}{1 - \mu r^2 (\gamma, \tilde{A}^{-1} \, \gamma)} \,\,.
\end{equation}
The function
\begin{equation} \label{densitychaplygin}
f(\gamma) := [1 - \mu r^2 (\gamma, \tilde{A}^{-1} \gamma)]^{-1/2} 
\end{equation}
 was found by
Chaplygin to  be   the density of an invariant measure in $\Re^6$:
\begin{equation}
\nu_{\Re^6} = f(\gamma) \, d\gamma_1 d\gamma_2 d\gamma_3 dL_1 dL_2 dL_3  
\end{equation} 
 This   follows from Veselov's theorem, as   $F(\gamma) = 1 - \mu r^2 (\gamma, \tilde{A}^{-1} \gamma)$ is (up to a constant factor) the determinant of the linear map $\Omega \mapsto L = L(\Omega; \gamma)$.  
 For direct proofs of  invariance of the measure, see \cite{Duistermaat} or \cite{Fedorov}. 

A system of ODE's for the rubber ball can be derived in a similar fashion.
For the angular momentum $\ell$ at the contact point, we get the same equation
(\ref{Veselova1}) from Veselova's system, but the relation between $\Omega$ and $L$ is (\ref{LOmega}),  the same as in Chaplygin's marbe. Differentiating
$(\Omega,\gamma) = 0$ the multiplier can be eliminated.  

\paragraph{Hamiltonization of Veselova's system.} The compressed Lagrangian  is 
\begin{equation} \label{Veselovalagrangian}
L_{comp} = \frac{1}{2} ( A(\dot{\gamma} \times \gamma), \dot{\gamma} \times \gamma) \,\,\,,
\end{equation}
 since  $ \Omega = \dot{\gamma} \times \gamma$;  the momentum map corresponding to the $S^1$-action is $J = \ell_3 = (L,\gamma)$.  Thus  ${\rm (J.K)} = \ell_3 \,d\rho_3 = (A \Omega, \gamma) \,d\rho_3$, where $d\rho_3$ is the area form of $S^2$. 
The compressed  Legendre transform is
$$ \dot{\gamma} \mapsto  a = \frac{\partial L^*}{\partial \dot{\gamma}} = \gamma \times  A(\dot{\gamma} \times \gamma) \,\,\,.
$$
The nonholonomic 2-form in $T^*S^2$ is
\begin{equation} \Omega_{NH} = da \wedge d\gamma + (A(\dot{\gamma} \times \gamma), \gamma)\,d\rho_3 
\end{equation}
Being a two-degrees of freedom system, a general result from \cite{Fedorov1} (Theorem 3.5) garantees that this system is Hamiltonizable. In order to verify  that $\Omega_{NH}$ is conformally symplectic, it is simpler to use $\dot{\gamma}$ themselves as coordinates, that is, we pull
back $\Omega_{NH}$ to $TS^2$ via $Leg^*$. We get
$$  \Omega_{NH} = d(\gamma \times A(\dot{\gamma} \times \gamma))) \wedge d\gamma + (\gamma, A(\dot{\gamma} \times \gamma))d\rho_3   \,\,\,.
$$
\begin{proposition} Veselova's system is conformally symplectic,
$d(f \Omega_{NH})= 0 $, with   conformal factor 
\begin{equation}
 f = f(\gamma) = (A^{-1} \gamma, \gamma)^{-1/2} \,\,.
\end{equation}
\end{proposition}
As expected, it is the density of the Veselovs invariant measure $\mu = f(\gamma) dL d\gamma$
obtained via Proposition \ref{Veselovstheorem}. The orthonormal frame in $S^2$ diagonalizing
(\ref{Veselovalagrangian}) provides explicit coordinates for integration via Hamilton-Jacobi's method.

\paragraph{Chaplygin's rubber ball.}  The dynamics compress  to $T^*S^2$, and by the same general result in \cite{Fedorov1}, we know in advance that
the system is Hamiltonizable. Choose a moving
frame $e_1, e_2$ in $S^2$.  The horizontal lift from $\dot{\gamma} = v_1\,e_1 + v_2\,e_2$ to
${\rm Hor}(\dot{\gamma}) \in T(SE(3))$ is easily done via (\ref{horSO3}):  
$$ {\rm Hor}(\dot{\gamma}) = v_2\, \left( X_1^r - r \partial/\partial y \right) - v_1\, \left( X_2^r + r \partial/\partial x \right)
$$
Composing $d\alpha_{rubber} = (\rho_1 \wedge \rho2, - r \rho_3 \wedge \rho_1, r \rho_2 \wedge \rho_3, 0)$ with ${\rm Hor}$, we get $K_{rubber} = (dS \hat{k},0,0,0)$, where $dS$ is the
$S^2$ area form. Thus for the term $(J.K)$  we need
only the third component of the angular momentum, $m_3 = (M,\gamma) = (A\Omega,\gamma)$, 
where we insert  (\ref{horSO3}) 
$\Omega = \dot{\gamma} \times \gamma = v_2 \,e_1 - v_1 \,e_2 $. Therefore
\begin{equation}
\Omega_{NH} = \Omega_{T^*S^2} + (A(\dot{\gamma} \times \gamma),\gamma) \cdot dS\,\,.
\end{equation}
Here $\dot{\gamma} = v_1\,e_1 + v_2\,e_2 \in TS^2$  corresponds to $p_{\gamma} = p_1 \theta_1 + p_2\,\theta_2$ via the Legendre map $Leg^{comp}$ of  the compressed Lagrangian
\begin{equation} 
L_{comp} = \frac{1}{2}  A (v_2\,e_1 - v_1\,e_2,v_2\,e_1 - v_1\,e_2) + \frac{1}{2} \mu r^2 (v_1^2 + v_2^2) \,\,\,.
\end{equation}
Clearly, this system becomes Veselova's for $r = 0$. Using Proposition \ref{Veselovstheorem}
and Fedorov's result for two degrees of system, we get:
\begin{proposition} \label{rubberdensity} The  compressed rubber ball   system is
Hamiltonizable.  The  conformal factor is 
\begin{equation}
f = [{\rm det} \, Leg^{comp}]^{- 1/2} = (I_1I_2I_3)^{-1/2}\, \left( (A^{-1} \gamma, \gamma) + \mu\,r^2 [ \frac{\gamma_2^2 + \gamma_3^3}{I_2I_3} + \frac{\gamma_1^2 + \gamma_3^3}{I_1I_3} + \frac{\gamma_1^2 + \gamma_2^3}{I_1I_2}] + \frac{\mu^2 r^4}{I_1I_2I_3} \right)^{-1/2} \,\,.
\end{equation} 
\end{proposition}
\noindent{\bf Proof.}\footnote{We can provide the (short) notebook under request. 
It should be investigated if the rubber ball problem is integrable. Does a (quartic) integral still exist? }  We checked using  spherical coordinates and {\it Mathematica{\small \copyright}}.

\subsection{Chaplygin's marble  is not Hamiltonizable at the $T^*SO(3)$ level.} 
\paragraph{The homogeneous sphere.} In a nutshell: the dynamics in the homogeneous case is embarrassingly simple. The angular velocity
in space is constant, so the attitude matrix $R$ evolves as a 1-parameter group
$ R = \exp([\omega] t)$, so $\Omega$ and $\omega$  are constant. The vector $\gamma(t)$ describes a circle in the sphere perpendicular to $\omega$,
and $L(t)$ the curve given by $ L(t) = (I + \mu r^2) \, \omega - \omega_3 \, \gamma(t) $.
Provided $\ell$ is   not vertical, $L$ and $\gamma$ are never parallel.
The invariant tori  are always foliated by closed curves and the two frequencies coincide.
From the constraint equations we see that the motion of the contact point in the plane is a straight line. Shooting pool with a perfect Chaplygin ball is very dull\footnote{We found the following relevant information in www.ot.com/skew/five/myths.html (Top Ten Myths in Pool
or the Laws of Physics Do Apply): 
``4. If the cue is kept level, contacting the cueball purely left or right of its center will make it curve as it rolls. ({\bf No!} The rolling cue ball can have two completely independent components to its angular momentum. Basically, this means that it can rotate in the manner of a top while rolling slowly forward along a straight line. In general, spin on a cue ball is of two types; follow/draw is the spin like tires on a car, while English is the spin like a child's toy 'top'. Seperately, neither one will make a ball curve! If they are combined - e.g., strike low-left giving left English and draw - then the spin is called masse (mass-ay), and the ball will curve as it travels.)''}.  {\it Let us use these simple results as template for  our
 operational system.}
In terms of the right coframe, we have   
\begin{equation}
\Omega_{NH} = d\ell_1 \rho_1 + d\ell_2 \rho_2 + d\ell_3 \rho_3 + \ell_1 \rho_2 \rho_3 + 
\ell_2 \rho_3 \rho_1 + \ell_3 \rho_1 \rho_2 - \mu r^2 (\omega_2 \rho_3 \rho_1 + \omega_1 \rho_2 \rho_3)
\end{equation}
{\it This formula holds in   general}.  In the nonhomogeneous case  one  must
write $\omega_1$ and $\omega_2$ in terms of   $\ell$  and $R \in SO(3)$:
$ \omega = R \Omega = R \Omega_{\gamma} ( R^{-1} \ell )
$
which seems to be  a quite involved expression, a haunting monster we will avoid, until a final
confrontation in Proposition \ref{marbleconformal}. In the homogeneous case, life is much easier:
$ \omega  =  R \frac{1}{\kappa} I R^{-1}  m  =  \frac{1}{I}  m\,,
$
so  the dependence of $\omega$ on $R$ disappears. 
The Hamiltonian is given by 
$$
  H = \frac{1}{2} \left( \frac{\ell_1^2 + \ell_2^2}{I + \mu r^2} \right) + \frac{\ell_3^2}{I}
$$
where
$$ \ell_1 = (1 + \frac{\mu r^2}{I}) m_1 \,,\,\ell_2 = (1 +  \frac{\mu r^2}{I}) m_2\,,\,\ell_3 = m_3 \,\,\,\,,\,\,\,\,
 \omega_1 = \frac{m_1}{I}\,, \,\omega_2 = \frac{m_2}{I}\,,\,\omega_3 = \frac{m_3}{I}\,.$$
To obtain the equations of motion we solve 
\begin{equation}
\left( \begin{array}{l} \omega_1 \\ \omega_2 \\ \omega_3 \\   \dot{\ell}_1 \\ \dot{\ell}_2 \\ \dot{\ell}_3 \end{array} \right) = 
\left( \begin{array}{llllll}  0 & 0 & 0 & 1 & 0 & 0 \\  0 & 0 & 0 &  0 & 1 &  0  \\  0 & 0 & 0 &  0 & 0 &  1 \\
 -1 & 0 & 0 &  0 & \ell_3 &  -I \omega_2 \\ 0  & -1 & 0 &  - \ell_3 & 0 & - I \omega_1 \\  0  &  0 & -1 &   I \omega_2 &   - I \omega_1 & 0 
\end{array} \right) \cdot  \left( \begin{array}{l} 0  \\ 0 \\ 0 \\   \omega_1 \\ \omega_2 \\ \omega_3 \end{array} \right)
\end{equation}
where we have used
$H_{\ell_1} = \ell_1/(I + \mu r^2) = m_1/I = \omega_1 \,\,,$ 
and similarly,
$ H_{\ell_2} =   \omega_2\,, H_{\ell_2} =   \omega_2 \,\,.$ 
This  gives, as expected:
$$  \dot{\ell}_1 = (I \omega_3) H_{\ell_2} - I \omega_2 H_{\ell_3} = 0\,,\, \dot{\ell}_2 = ... = 0\,,\,\, \dot{\ell}_3 = ... = 0  \,\,\,.
$$
Thus  
$  \omega_i = m_i/I = {\rm const} \,\,\,,\,\,  i = 1,2,3 \,\, $
and the vectorfield is simply 
$ X = \omega_1 X_1^{\rm right} + \omega_2 X_2^{\rm right} + \omega_3 X_3^{\rm right} \,\,$
(no components in the fiber directions $\partial/\partial m_i$).
We  now use Theorem
\ref{criterionHamiltonizable}. Using $m$ as coordinates, the nonholonomic two form is given by
$$ \Omega_{NH} = (1 + \frac{\mu r^2}{I})\,(dm_1 \rho_1 + dm_2 \rho_2) +  dm_3 \rho_3 +  ( m_1 \rho_2 \rho_3 + m_2 \rho_3 \rho_1 + m_3 \rho_1 \rho_2 )
$$ 
so that
$d\Omega_{NH} = - \frac{\mu r^2}{I} (dm_1 \rho_2 \rho_3 + dm_2 \rho_3 \rho_1) \,\,.
$
It is easy to see that the equation $d\Omega_{NH} = \Omega_{NH} \wedge \alpha $
has no solution whatsoever.  Indeed, suppose
 $\alpha = A_1 dm_1 + A_2 dm_2 + A_3 dm_3 + B_1 \rho_1 + B_2 \rho_2 + B_3 \rho_3 \,\,.$  Taking the exterior product, and looking at terms like $dm_1 dm_2 \rho_2$
we see that all the $A$'s must be zero.  Examining the coefficient of $ \rho_1 \rho_2 \rho_3$
we get $B_1 m_1 + B_2 m_2 + B_3 m_3 \equiv 0$ so all the $B$'s are also zero.

Hence the homogeneous Chaplygin sphere,  as simple at it can be, has no conformal symplectic structure! In fact, it does not have an affine symplectic structure either.
A short calculation shows that
$$  i_X d\Omega_{NH} = \frac{\mu r^2}{I^2} ( - dm_1 m_2 \rho_3 + dm_1 \rho_2 m_3 - dm_2 m_3 \rho_1 + dm_2 \rho_3 m_1) \neq 0 .$$ 
By continuity,  for sufficiently close but different inertia coefficients the inequalities persists. We have also done the calculation for the non homogeneous case and things only get worse.
But, it still remains a possibility: is the {\it reduced} system to $T^*S^2$  Hamiltonizable?
The impatient reader can go directly to Theorem \ref{marbleconformal}.

\subsection{Chaplygin's marble: reduction  to $T^*S^2$ } 

Using (\ref{Lb}), $L = a \times \gamma + \ell_3\, \gamma$, Chaplygin's marble equations in $(L,\gamma)$-space directly reduce to $T^*S^2$:
\begin{equation} \label{reducedChaplygin}
 \dot{\gamma} = \gamma \times \Omega\,\,,\,\, \dot{a} =  - 2H \, \gamma + (\gamma,\Omega) \cdot(a \times \gamma + \ell_3 \, \gamma) 
\end{equation}
with
$$\Omega = \Omega(a,\gamma; \ell_3) =  \tilde{A}^{-1}L + \mu r^2 \, \frac{(\gamma, \tilde{A}^{-1} L)}{1 - \mu r^2 (\gamma, \tilde{A}^{-1} \, \gamma)} \, \tilde{A}^{-1}(\gamma) \,\,.
$$

\paragraph{$S^1$ reduction of the homogeneous sphere to $T^*S^2$.}  
The homogeneous Chaplygin sphere  
when reduced to $T^*S^2\,,$  produces a more interesting system. Equations (\ref{reducedChaplygin}) become 
\begin{equation} \label{a}
\dot{\gamma} = \frac{1}{I + \mu r^2} \, a\,\,\,\,\,,\,\,\,\,\, \dot{a} = \omega_3 \, a \times \gamma -  \frac{1}{I + \mu r^2}\,|a|^2 \gamma
\end{equation}
One observes  that $(a,\gamma) = 0$ and that $|a|^2$ is conserved. So at each level set,
we get an isotropic 3d-oscillator with a  Lorentz force\footnote{Alan Weinstein commented in more than one occasion that ``unreduction''  
sometimes is even nicer than reduction: unreducing a non-trivial system may lead  to an
trivial one. Alan credits this moto to Guillemin and
Sternberg;   one
reference could be \cite{Guillemin}.}.

\paragraph{Dimension count argument.} In hindsight, we can give   two simple arguments why the Chaplygin marble could not   be
Hamiltonizable at  the $T^*SO(3)$ level.   First:
if $(T^*SO(3),\Omega_{NH}, H)$  were Hamiltonizable, the system would be Liouville integrable
 {\it by ``mere'' symmetries}, due to
the existence of three independent first integrals $H, \ell_3, \ell_1^2 + \ell_2^2, \ell_3^2$.
But it is known that  integrability of Chaplygin's marble stems not from symmetries, but from a special choice of separating coordinates (\cite{Duistermaat}), namely, elliptic coordinates of the sphere.  Second:  \cite{Stanchenko} verified that 
  Chaplygin's density function  F (\ref{densitychaplygin}) of the system in $\Re^6$ also gives   an invariant measure on $T^*SO(3)$ (see also \cite{Duistermaat}, section 7),  
\begin{equation}
\nu_{T^*SO(3)} = F(\gamma) d\lambda_1 d\lambda_2 d\lambda_3 dL_1 dL_2 dL_3 \,\,,\,\,\, F = [ 1 - \mu r^2 (\gamma, \tilde{A}^{-1} \gamma) ]^{-1/2}.
\end{equation}
 Were the compressed system Hamiltonizable in
$T^*SO(3)$, the conformal factor (time reparametrization) should be $ F(\gamma)^{\frac{1}{m-1}}$, with $m=3$, see (\ref{dimensioncount}).
But  the  correct time reparametrization holds with $m=2$ instead of $m=3$.  This strongly suggests
that Hamiltonization should be attempted {\it after} reduction of the internal $S^1$
symmetry.

\paragraph{Phase locking.} The fact that Chaplygin's sphere is integrable implies an interesting {\it phase locking} property. For simplicity, consider a resonant torus and a periodic solution,   $\gamma(T) = \gamma(0)\,,\,L(T) = L(0)$.  We may assume that $R(0) = {\rm identity}$ so $R(T)$ preserves both $k$ and $\ell$.   If we assume $\ell \neq \pm k$, then $R(T) $ must   also  be the identity (there is only one orthogonal matrix with two different eigenvectors with equal eigenvalues 1). Since the rotational conditions are reproduced after time $T$, there is a ``planar geometric phase'' (meaning a translation), $\Delta z = (\Delta x, \Delta y)$.   From \cite{Duistermaat},
section 11, one knows this direction:
\begin{proposition} In average, $\Delta z$ moves in the direction of $\ell \times k$.
\end{proposition}
In the normal direction $k \times (\ell \times k)$ there is a ``swaying motion'', with
zero average, see \cite{Duistermaat}, (11.71), and remark 11.11. This  result depends on the
explicit solution in terms of elliptic coordinates, but the zero average can be proved
in a more elementary way, see Duistermaat, section 8.2.  In the direction $\ell \times k$ one has
$$ \frac{d}{dt}\, (z(t), \ell \times k) = r\, (\omega \times k\,,\,\ell \times k) = r\, (\omega\,,\,\ell - \ell_3 \,k) = r\, \left( 2T - \ell_3 \omega_3\,\right) > 0 \,\,.
$$
\cite{Duistermaat} shows (section 9.2) that by a suitable change of coordinates, one may assume
that $\ell_3 = 0$, so in this equivalent problem, the velocity in this direction
is simply $2rT$. 

\paragraph{Chaplygin's marble  via the almost Hamiltonian structure.} After this detour, we hope the reader will   appreciate a concise way to describe this system.
The clockwise map is 
\begin{equation} \begin{array}{lllllll}  \label{diagram2} 
 T(SO(3) \times \Re^2) & \longrightarrow & T^*(SO(3) \times \Re^2) &  \,\,\,\,\,  (\Omega, \dot{x}, \dot{y}) & \mapsto  & (M = A\Omega, P_x = \mu \dot{x}, P_y = \mu \dot{y}) & \\
& Leg &  & & & & \\
\, \uparrow & \mbox{       }& \,\,   & \,\,\,\,\,\,\,\,\,\,\,\,\,\,\,\,\,\,\,\uparrow   & & &         \\
\,   h  &   & \mbox{ }& \,\,\,\,\,\,\,\,\,\,\,\,\,\,\,\,\,\,\,h & & & \\
\,\,  | &         & \,\, & \mbox{              } \,\,\,\,\,\,\,\,\,\,\,\,\,\,\,\,\,\, | &   &  \\
& &  &  & &  &\\
TSO(3) & \longleftarrow & T^*SO(3) &   \,\,\,\,\,\,\,\,\,\,\,\,\,\,\,\,\,\,\,  \Omega & \leftarrow & \,\,\, L & \\
& (Leg^{\phi})^{-1} &
\end{array}
\end{equation}
where
$(\dot{x},\dot{y}) = r \omega \times k \,\,$, and $ (P_x,P_y) = \mu r \omega \times k $ . 
We now compute the ``gyroscopic'' 2-form:
\begin{equation} \label{g}
 {\rm (J.K)} =   r ( -  P_x d \rho_2 + P_y d\rho_1 ) =   \mu r ( -  \dot{x} d \rho_2 + \dot{y} d\rho_1 ) =  - \mu r^2 ( \omega_2 d \rho_2 + \omega_1 d\rho_1 )
\end{equation}
To obtain $\omega_1$ and $\omega_2$ as  functions in $T^*SO(3)$,  we use the Legendre transformation:   
$\omega = R \Omega = R A^{-1} M $
so  ${\rm (J.K)} $  is a combination of the basic forms $\rho_3  \wedge \rho_1\,  , \,   \rho_2 \wedge \rho_3$ (coefficients  linear in $M$ and   functions of $R$).

\paragraph{$\,\,S^1$ invariance.} We claim that $\Omega_{NH}$ is $S^1$-invariant    ($S^1$ acting only in the first factor of $Q = SO(3) \times \Re^3$). For the canonical term this is a standard symplectic fact. The $(J.K)$ term is invariant as well: 
 (\ref{g}), written in terms of the  left-invariant forms,  depends only on the Poisson vector $\gamma$:
\begin{equation} \label{leftjk}
 {\bf {\rm (J.K)}} = \mu r^2 \left( \gamma \times (\Omega(L,\gamma) \times \gamma) \,,\, d \lambda \right) 
\end{equation}
In fact,  the $S^1$ action  generated by
the right invariant vectorfield  $X_3^{\rm right}$ 
  maintains the projection $\gamma$ fixed.  We know  (general nonsense) that the right invariant vectorfields preserve the left invariant forms:
$ R_{\phi}^* \, \lambda_i  = \lambda_i \,\,\,.$  
(Proof: $(R_{\phi}^* \, \lambda_i)(\dot{R}) = \lambda_i (R_{\phi} R [\Omega])
= \Omega_i $). Since under the left $S^1$ action (actually under the left action of $SO(3) $ on $SO(3)$) the {\it value} of $\Omega$ remains unchanged,
the ${\rm (J.K)}$ term is preserved.

\paragraph{The twisted action generator and $S^1$ reduction.}  MW reduction method works fine,
although $X_3^r $
is the  Hamiltonian vectorfield  of $ J = \ell_3 $, relative
to the canonical symplectic form, but not relative to $\Omega_{NH}$. 
We just change to the {\it twisted } $S^1$-action generator
$\tilde{X}_3$, defined by $i_{\tilde{X}_3}\,\Omega_{NH} = - d\ell_3$.  A simple computation gives 
$
\tilde{X}_3 = X_3^r - m_2\, \frac{\partial}{\partial \ell_1} + m_1\, \frac{\partial}{\partial \ell_2}
$
where $m_1 = \ell_1 - \mu r^2 \omega_1\,\,,\,\, m_2 = \ell_2 - \mu r^2 \omega_2 $.
 The reduced manifold is the quotient
of a level $\ell_3$ in $T^*SO(3)$, identifying the flow lines $\tilde{\phi}$. A concrete realization is achieved using  (\ref{sectioni}). Taking the pullback via $i^*$, the reduced form is then
\begin{equation} \label{reducednh}
\Omega_{red} = \Omega^{can}_{T^*S^2} + \ell_3 \,{\rm area}_{S^2} - \mu r^2 (\omega_1\, d\theta_2 - \omega_2 \, d\theta_1)
\end{equation}
where we recall the parametrizations $p_1 \theta_1 + p_2 \theta_2 \in T^*S^2\,,\, R(\gamma) =
{\rm rows}(e_1,e_2,\gamma) $. 

In (\ref{reducednh}) we must write $\omega_1, \,\omega_2$ 
explicitly in terms of $p_1,\,p_2, \, \ell_3$.  To write this explicitly, there is no other option than to confront
the monster (which actually is not that terrible):
from $\Omega = R^{-1} \omega = \omega_1\,e_1 + \omega_2\,e_2 + \omega_3\,\gamma$ and from
(\ref{LOmega}) we get
\begin{equation} \label{lomega}
\omega =  R(\gamma) \,\Omega_{\gamma}\, [R(\gamma)]^{\dagger}  \, \ell   \,\,\,,\,\,\, \ell = (p_2, - p_1, \ell_3) \,\,. 
\end{equation}
where  $\Omega_{\gamma} $ is explicitly given by (\ref{OmegaL}).

\begin{theorem} \label{marbleconformal}
$ i_X \,d( f \Omega_{red}) \neq 0 \, \,, \,\,\, f(\gamma) = [1 - \mu r^2 (\gamma, \tilde{A}^{-1} \gamma)]^{-1/2} \,\,.$
\end{theorem}
\noindent {\bf Proof.} We used  spherical coordinates ({\it f\^aute de mieux}) and a {\it Mathematica{\small \copyright}} notebook. It misses being conformally symplectic by very little (even in the homogeneous case)\footnote{\cite{Borisovsphere1}   showed by a subtle numerical evidence that,
in the
original time, Chaplygin's marble is not Hamiltonizable  at any level of reduction. The question whether Chaplygin's marble is Hamiltonizable in the new time $ dt/d\tau = f(\gamma)$ 
was addressed in  \cite{Borisovsphere}.  
 They provide a bracket structure in terms of the coordinates $(L,\gamma)$ or the coordinates $(\tilde{L},\gamma)$, with $\tilde{L} = L/f(\gamma)$. Using a computer algebra program we
checked that the second brackets satisfy the Jacobi identity. However, we could not recover
Chaplygin's equations for the $L$ coordinates, even in the homogenous case.}.  

Our calculation suggests that Chaplygin's sphere is {\it not} affine symplectic even
at the $T^*S^2$ level, so  Chaplygin's sphere integrability is due to a specific
nonholonomic phenomenon.  This observation is in accordance with the opening statement in \cite{Duistermaat}: 
\begin{quote}``Although the system
is integrable in every sense of the word, it neither arises as a Hamiltonian system, nor is the integrability an immediate consequence
of the symmetries''.
\end{quote}

\section{Recent developments and final comments}

NH systems   have a reputation of  having peculiar (even rebellious) dynamic behaviour  (\cite{Arnold2}). In spite of good progress, the general theory for NH systems is  way behind the theory of Hamiltonian systems. For instance, although the groundwork for
an Hamilton-Jacobi theory for NH systems has been set up in \cite{Weber}, not much has
been achieved since then. 

We have no intent (nor competence) to make a  survey of  recent developments in NH systems, specially regarding reduction of symmetries; nevertheless  it may be worth  registering the intense  activity going on.  Recent books of interest  are  \cite{Cushmanbook},  \cite{Cortes1}, \cite{Oliva}, \cite{Bloch} and a   treatise in the Mechanical Engineering tradition is \cite{Papastavridis}\footnote{Reviewed in \cite{Koillerreview}.}.  {\em Reports on Mathematical Physics} has been  publishing NH papers regularly, and {\em Regular and Chaotic Dynamics} devoted large parts of vols. 1/2 (2002) to NH systems. For older eastern European literature  see   {\em PMM USSR, J. Applied Math. and Mechanics}, which has strongly influenced Chinese
mechanicists as well.  For a historical account on NH systems, from a somewhat ``anti-reducionist'' perspective, see \cite{Borisov}.

\subsection{Invariant measures and integrability}
 \cite{KupkaOliva} and \cite{KobayashiOliva}  find   conditions  insuring a special, but
 very interesting situation, where the Riemann  measure in $TQ$ induced by the metric   in Q,  is an invariant measure 
 for the NH system. Invariant measures for systems
with distributional symmetries were characterized in \cite{Blochnonlinearity}.

Curiously, although  a number of interesting NH systems have been  solved  using, say,  Abelian functions, a precise definition for integrability of a NH system is still lacking (\cite{BatesCushman}). 
These examples suggest  that the presence of an invariant measure must be imposed as a necessary (although not   sufficient) condition for integrability (whatever it may be), see \cite{Kozlov}. 
Most of them have enough integrals of motion   that the dynamics occurs on invariant 2-dimensional tori. Due to  the invariant measure, {\it the flow becomes linear in these tori after a time rescaling}. This follows from Jacobi's  multiplier method and
 Kolmogorov's theorem   (\cite{Arnold1}).
   Time reparametrization indicates the possibility of an affine symplectic structure. We believe that characterizing   NH systems possessing an affine symplectic structure (if needed, after some reduction stage) could be an interesting project. As a first
step, one may examine   the exisying literature to see which examples fit.
We list a few papers for that purpose: \cite{Veselovs},
\cite{Veselovs1}, \cite{Fedorov2}, \cite{Cushmanetala95},  \cite{Zenkov}, \cite{ZenkovBloch}, \cite{Dragovic},
\cite{Jovanovic}, \cite{Fedorov1}. One can hope that the manifestly geometric character of (\ref{compressed})
 can be instrumental to understand   when, where and why {\it Hamiltonization} 
is possible. Moreover, a prior geometric understanding of the invariant volume form
conditions is a more general question. It would be also interesting to tie the ``Hamiltonizable'' question
with the   invariants from the Cartan equivalence viewpoint, see below.

\subsection{Nonholonomic reduction.} The difficulties in  reduction  for general NH systems are explained in \cite{Sniatycki2002}. There are
four current theories of reduction of symmety for nonholonomic systems\footnote{We thank one of the referees for this information.}:
i) {\it projection methods}, see \cite{Marle},  \cite{Dazord};
ii) {\it the distributional Hamiltonian approach}, initiated by \cite{Bocharov}
and developed in \cite{BatesSniatycki}, \cite{Cushmanetal95}, \cite{Sniatycki98}, and \cite{CushmanSniatycki}.
iii) {\it bracket methods}, initiated by \cite{MS},  and developed by \cite{Koon} 
and \cite{Sniatyckisingular};
iv) {\it Lagrangian reduction}, see \cite{Cemara}.

A few other references in this rapidly developing theme, besides those already mentioned  are: \cite{Bates},   \cite{Koon1}, \cite{Cantrijn1}, \cite{Cortes}, \cite{Marle1},
\cite{Marle2}.

\paragraph{Almost Poisson, almost Dirac approaches.}
  \cite{MS} were the first to describe a NH system   using an
almost-Poisson structure\footnote{Physicists are never
shy to use the word ``super'' in their endeavours;  on the other hand we, mathematicians, prefer
to use  low key terminology, like ``almost-quasi-twisted-(freakaz-)-`oid's''; this certainly does not help our image problem
with applied people, see \cite{Papastavridis} and \cite{Koillerreview}.},  
$ \dot{x_i} = \{x_i, H \}_{MS} $.
This bracket, defined  on the manifold $P = Leg({\cal H}) \subset T^*Q$,
where $Leg: TQ \rightarrow T^*Q$ is the Legendre transformation, in general does not
satisfy the Jacobi identity. They proved that  the Jacobi identity holds if
and only if the constraints are integrable.  
In \cite{Koillerromp} we gave a moving frames based derivation of the bracket structure. 
For some recent work on the MS-bracket and also Dirac estructures (the latter introduced in
\cite{Courant}), see \cite{Cantrijn3}, \cite{Koon},
\cite{Ibort}, \cite{Gallardo}. 
In spite of these advances, a complete understanding of the NH bracket geometry is still in order\footnote{Observations by J. Marsden (joint work with H. Yoshimura),
and by C.Marle, in their Alanfest talks, are important steps in this direction.}.

\subsection{$G$-Chaplygin systems via affine connections} 
Trajectories of the compressed  system can be described as geodesics of an affine connection  $\nabla^{NH}$ in $S$
(\cite{VershikFadeev}, \cite{KoillerARMA}).
For background in this approach, see \cite{Lewis} and references
therein.  Consider the parallel transport
operator along closed curves; if the  holonomy group is always conjugate to a subgroup of $S0(m)$, then
the connection is metrizable.  This means that there is a metric such that $\nabla^{NH}$ is precisely the Levi-Civita connection of this metric.
More generally, one may want to know when the geodesics of $\nabla^{NH}$ are, up to time reparametrization, the geodesics of a Riemannian metric. This is a traditional area in differential geometry, whose roots go back to the 19th century, and  goes under the name
of {\it projectively equivalent connections} (\cite{Cartan4}, \cite{Eisenhart}, \cite{Kobayashi}, \cite{Sharpe}). \cite{Grossman}   studies integrability of geodesics
equations via the equivalence method.
Our problem, then, is to find conditions for the NH connection to be projectively equivalent to a Riemannian connection. 
It would be also interesting to tie the  Hamiltonization  question
with the  
canonical system and invariants of the Cartan equivalence method. When an internal
symmetry group   is present, it would be desirable to construct a projected connection
in $S$ for each set of conserved momenta, and address these 
issues in the reduced level.

\bigskip

\noindent {\bf Acknowledgements.}  
Thanks to Hans Duistermaat for informations on  
Chaplygin's sphere and the referees for very good criticims and suggestions.
This is also a special occasion   to thank   our mathematical family: 
Alan, of course;  mathematical brothers and sisters from all continents (specially Yilmaz Akyildyz and Henrique Bursztyn) and cousins (specially Tudor Ratiu and Debra Lewis), ``uncle'' Jerry Marsden and, last but not least,  ``grandfather''  Chern (with his
gentle   voice, commanding us to keep interested in Math, up to his age).  

\smallskip
\centerline{We shall toast  in many more Alanfests: {\it Madadayo!}}
\bigskip



{\footnotesize \begin{thebibliography}{99}

\bibitem[Abraham and Marsden(1994)]{AM}
Abraham, R.,  Marsden, J. [1994], {\em Foundations of Mechanics},
\newblock  Perseus Publishing, 2nd revised  edition, 1994. 

\bibitem[Arnold(1989)]{Arnold1}
Arnold, V.~I. [1989], {\em Mathematical Methods of Classical Mechanics},
volume~60 of {\em Graduate Texts in Math.}
\newblock Springer-Verlag, First Edition 1978, Second Edition, 1989.

\bibitem[Arnold, Kozlov, and Neishtadt(1988)]{Arnold2}
Arnold, V.~I.,  Kozlov, V.V.,  A.~I. Neishtadt, A.I. [1988],
{\em Mathematical aspects of classical and celestial mechanics},
in Arnold, V.~I., editor, {\em Dynamical Systems III}.
\newblock Springer-Verlag, 1988.

\bibitem[Bates(2002)]{Bates}
Bates, L. [2002], Problems and progress in nonholonomic reduction, {\em Rep. Math. Phys. } \textbf{49}:2/3, 143--149.

\bibitem[Bates and Cushman(1999)]{BatesCushman} 
Bates, L., Cushman, R. [1999], What is a completely integrable nonholonomic dynamical system?,
{\em  Rep. Math. Phys. } \textbf{44}:1-2, 29--35. 

\bibitem[Bates and \'Sniatycki(1993)]{BatesSniatycki}
Bates, L., \'Sniatycki, J. [1993], Nonholonomic Reduction, {\em Rep. Math. Phys.} \textbf{32}:1, 99--115.

\bibitem[Blackall(1941)]{Blackall}
Blackall, C.J. [1941], On volume integral invariants of non-holonomic dynamical systems,
{\em Amer. J. Math.} \textbf{63}:1, 155--168.

\bibitem[Bloch(2003)]{Bloch}
Bloch, A.M. [2003], {\em  Nonholonomic mechanics and control},
\newblock  Springer-Verlag, 2003.

\bibitem[Bloch, Krishnaprasad, Marsden and Murray(1996)]{BKMM}
Bloch, A.M., Krishnaprasad, P.S., Marsden, J.E., Murray, R.M. [1996], Nonholonomic mechanical
systems with symmetry, {\em Arch. Ratl. Mech. Anal.}, \textbf{136}, 21--99.

\bibitem[Bocharov and Vinogradov(1977)]{Bocharov}
A.V. Bocharov, A.V., Vinogradov, A.M.[1977], The Hamiltonian form of mechanics
with friction, nonholonomic mechanics, invariant mechanics, the theory
of refraction and impact, appendix II in: ``The structure of Hamiltonian
mechanics'', B.A. Kuperschmidt and A.M. Vinogradov (authors), {\em Russ.
Math. Surv.}, \textbf{42}:2, 177--243.

\bibitem[Borisov and Mamaev(2001)]{Borisovsphere}
Borisov, A. V., Mamaev, I.S. [2001], Chaplygin's ball rolling problem is Hamiltonian,
{\em Mathematical Notes (Matematicheskie Zametki)}, \textbf{70}:5, 793-795.

\bibitem[Borisov and Mamaev(2002)]{Borisovsphere1}
Borisov, A. V., Mamaev, I.S. [2002], Obstacle to the Reduction of Nonholonomic Systems
to the Hamiltonian Form, {\em Doklady Physics USSR}, \textbf{47}:12, 892--894.

\bibitem[Borisov and Mamaev(2002a)]{Borisov}
Borisov, A. V., Mamaev, I. S. [2002a], On the history of the development
of the nonholonomic dynamics,   {\em Regular and Chaotic Dyn.} \textbf{7}:1, 43--47.

\bibitem[Borisov and Mamaev(2002b)]{Borisov1}
Borisov, A. V., Mamaev, I. S. [2002a],
The Rolling Body Motion Of a Rigid Body on a Plane and a 
Sphere. Hierarchy of Dynamics,
{\em Regular and Chaotic Dyn.}, \textbf{7}:2, 177--200.

\bibitem[Borisov, Mamaev and Kilin(2002)]{Borisov2}
Borisov A.V., Mamaev I.S. and Kilin A.A.[2002], The rolling motion of a ball on
a surface. New integrals and hierarchy of dynamics, {\em Regular and Chaotic 
Dyn.}, \textbf{7}:2, 201–-218.

\bibitem[Bryant(1994)]{Bryant1} 
Bryant, R. [1994], Lectures on the Geometry of Differential Equations, (The Fall 1994 André Aisenstadt Lectures given at the Centre de Recherche Mathematique in Montreal), in preparation.
See unpublished lecture notes, available at {\it http://www.cimat.mx/ ~gil}

\bibitem[Cantrijn, de L\'eon, Marrero and de Diego(1998)]{Cantrijn1}
Cantrijn, F., de L\'eon, M., Marrero, J.C., de Diego, D. [1998], 
Reduction of nonholonomic mechanical systems with symmetries,
{\em Rep. Math. Phys.} \textbf{42}:1/2, 25--45.

\bibitem[Cantrijn, de L\'eon, de Diego(1999)]{Cantrijn3}
Cantrijn, F.,  de L\'eon, M., de Diego, D. [1999], On almost-Poisson structures in nonholonomic
mechanics, {\em Nonlinearity} \textbf{12}, 721--737.

\bibitem[Cantrijn, Cort\'es, de L\'eon, and de Diego(2002)]{Cantrijn}
Cantrijn, F., Cort\'es, J., de L\'eon, M., de Diego, D. [2002], On the geometry of generalized
Chaplygin systems, {\em Math. Proc. Camb. Phil. Soc.} \textbf{132}, 323--351.

\bibitem[Cartan(1910)]{Cartan3}  
Cartan, \'E. [1910], Le syst\`emes de Pfaff \`a cinq variables et les \'equations aux d\`eriv\`ees partielles du second ordre,  {\em Ann. Sci. \`Ecole Normale } \textbf{27}:(3), 109--192.

\bibitem[Cartan(1926)]{Cartan0} 
Cartan, \'E. [1926], {\em La th\'eorie des groupes finis et continus et la g\'eom\'etrie diff\'erentielle, trait\'ees par la m\'ethode du repere mobile. Leçons profess\'ees à la Sorbonne} (R\'edig\'ees par Jean Leray), 
\newblock  Gauthier-Villars, Paris, 1951. 

\bibitem[Cartan(1928)]{Cartan2} 
Cartan, \'E. [1928], Sur la repres\'entation
g\'eom\'etrique
des syst\`emes mat\'eriels non holonomes, {\em Proc. Int. Congr. Math. Bologna} \textbf{4},    253--261.

\bibitem[Cartan(1937)]{Cartan4} Cartan, \'E [1937], {\em Le\c cons sur la th\'eorie des
espaces a connexion projective},
\newblock Gauthier-Villars, Paris, 1937.

\bibitem[Cartan(2001)]{Cartan1} 
Cartan, \'E. [2001], {\em Riemannian geometry in an orthogonal frame. From lectures delivered by \'Elie Cartan at the Sorbonne in 1926--27. With a preface to the Russian edition by S. P. Finikov. Translated from the 1960 Russian edition by Vladislav V. Goldberg and with a foreword by S. S. Chern},  
\newblock World Scientific Publishing Co., Inc., River Edge, NJ, 2001. 

\bibitem[Cendra, Lacomba and Reartes(2001)]{Celare}
Cendra, H., Lacomba, E.A., Reartes, W. [2001], 
The Lagrange-d'Alembert-Poincar\'e equations for the symmetric rolling sphere, {\em
Proc.  Sixth Dr. Antonio A. R. Monteiro Congress of Math. (Bah\'{\i}a Blanca, 2001)}, 19--32, 
Univ. Nac. Sur Dep. Mat. Inst. Mat., Bah\'{\i}a Blanca, 2001. MR1919465 (2003g:70021)

\bibitem[Cendra, Marsden and Ratiu(2001)]{Cemara}
Cendra, H., Marsden, J. E., Ratiu,  T. S. [2001], Geometric mechanics, Lagrangian reduction and nonholonomic systems, {\em Mathematics Unlimited-2001 and Beyond}, (B. Enguist and W. Schmid, eds.), Springer-Verlag, New York, 221--273. 

\bibitem[Cendra, Ibort, de L\'eon, de Diego (2004)]{Manoloprep}
Cendra, H.,  Ibort, A.,  de L\'eon, M., de Diego, D. [2004] 
A Generalization of Chetaev's Principle for a Class
of Higher Order Non-holonomic Constraints, submitted to J. Math. Phys.

\bibitem[Chaplygin(1911)]{Chaplygin0}
Chaplygin, S.A. [1911], On the theory of the motion of nonholonomic systems. Theorem
on the reducing factor, {\em Mat. Sbornik}, \textbf{28}, 303--314. 

\bibitem[Chaplygin(1981)]{Chaplygin} 
Chaplygin, S.A. [1981] {\em Selected works on Mechanics and Mathematics}
\newblock   Nauka, Moscow, 1981.

\bibitem[Chaplygin(2002)]{Chaplygin1}
Chaplygin, S.A. [2002], On a ball's rolling on a horizontal plane, {\em Reg. Chaot. Dyn.},
\textbf{7}:2, 131--148. \\  Original paper in {\em Math. Sbornik} \textbf{24}, 139--168, (1903).

\bibitem[Chern et.~al.(1991) Bryant, Chern,  Gardner, Goldschmidt, and Griffiths]{Chern} 
Bryant, R.L., Chern, S.S., Gardner, R.B., Goldschmidt, H.L, and Griffiths, P.A. [1991], {\em Exterior Differential Systems}, volume 18 of {\em MSRI Publications}, 
\newblock Springer-Verlag, 1991.

\bibitem[Clemente-Gallardo, Maschke and van der Schaft(2001)]{Gallardo}
Clemente-Gallardo, J., Mashke, B., van der Schaft, A.J. [2001], Kinematical constraints
and algebroids, {\em Rep. Math. Phys.} \textbf{47}:3, 413--429.

\bibitem[Cort\'es(2002)]{Cortes1}
Cort\'es, J[2002], {\em Geometric, Control and Numerical Aspects of Nonholonomic Systems},
 \newblock Springer-Verlag, 2002. 

\bibitem[Cort\'es and de L\'eon(1999)]{Cortes}
Cort\'es, J., de L\'eon, M.[1999], Reduction and reconstruction of the dynamics of nonholonomic systems, {\em J. Phys. A: Math. Gen.} \textbf{32}, 8615--8645.

\bibitem[Cort\'es, de L\'eon, de Diego and Mart\'{\i}nez (2003)]{CortesManolo}
Cort\'es, J., de L\'eon, M., de Diego, D., Mart\'{\i}nez, S.[2003], Geometric description of vakonomic and nonholonomic dynamics; comparison of solutions,
{\em SIAM J. Control Optim.} \textbf{41}:5, 1389--1412.

\bibitem[Courant(1990)]{Courant} 
Courant, T.J. [1990], Dirac Manifolds, {\em Trans. Am. Math. Soc.} \textbf{319}:2, 631--661.

\bibitem[Cushman, Hermans and Kemppainen(1995)]{Cushmanetala95}
Cushman, R., Hermans, J., Kemppainen,D.[1995], The rolling disk, in {\em Nonlinear dynamical
systems and chaos (Groningen, 1995)}, Prog. Nonlin. Diff. Eqs. Appl. \textbf{19}, 21--60, Birkh\"auser, Basel.

\bibitem[Cushman, Kemppainen, \'Sniatycki and Bates(1995)]{Cushmanetal95}
Cushman, R., Kemppainen, D., \'Sniatycki, J., Bates, L. [1995], Geometry of nonholonomic
constraints, {\em Rep. Math. Phys.} \textbf{36}:2/3, 275--286.

\bibitem[Cushman and Bates(1997)]{Cushmanbook}
Cushman, R., Bates, L. [1997], {\em Global aspects of Classical Integrable Systems},
\newblock Birkh\"auser, Basel, 1997.

\bibitem[Cushman and \'Sniatycki(2002)]{CushmanSniatycki}
Cushman, R., \'Sniatycki, J. [2002], Nonholonomic reduction for free and proper actions, 
{\em Reg. Chaotic Dyn.}, \textbf{7}, 61--72.

\bibitem[Dazord (1994)]{Dazord}
Dazord, P. [1994], M´ecanique hamiltonienne en presence de contraintes, {\em Illinois
J. Math.}, \textbf{38}, 148–-175.

\bibitem[Dragovic, Gajic and Jovanovic(1998)]{Dragovic}
Dragovi\'c, V., Gaji\'c, B., Jovanovi\'c, B.[1998], Generalizations of classical integrable nonholonomic rigid body systems, {\em J. Phys. A } \textbf{31}:49, 9861--9869. 

\bibitem[Duistermaat(2000)]{Duistermaat} 
Duistermaat, J.J. [2000], Chapygin's sphere,   in R. Cushman, J. J. Duistermaat and J. \'Sniatycki: {\em Chaplygin and the Geometry of
Nonholonomically Constrained Systems} (in preparation). 

\bibitem[Eisenhart(1925)]{Eisenhart} 
Eisenhart, L.P. [1925], {\em Riemannian Geometry},
\newblock Princeton University Press, Princeton, fifth printing, 1964.

\bibitem[Ehlers(2002)]{Ehlers}  
Ehlers, K. [2002],  The geometry of nonholonomic three-manifolds, {\em  in Proc.  Fourth Int. Conf. on Dynamical Systems and Differential Equations} 
\newblock Wilmington, NC, May 2002.

\bibitem[Fedorov(1989)]{Fedorov2}
 Fedorov, Yu. N.[1989], Two integrable nonholonomic systems in classical dynamics. (Russian), {\em Vestnik Moskov. Univ. Ser. I Mat. Mekh.} \textbf{4}, 38--41.

\bibitem[Fedorov and Kozlov(1995)]{Fedorov} 
Fedorov, Yu. N., Kozlov, V.V. [1995], Various aspects of n-dimensional rigid body dynamics, in Kozlov, V.V. (editor) {\em Dynamical Systems in
Classical Mechanics}, volume 168 of {\em AMS Translations series 2}, 1995.

\bibitem[Fedorov and Jovanovic(2003)]{Fedorov1}
Fedorov, Yu. N.,  Jovanovic, B. [2003],
Nonholonomic LR systems as Generalized Chaplygin systems with an Invariant Measure and Geodesic Flows on Homogeneous Spaces,
arxiv.org/abs/math-ph/0307016.

\bibitem[Flanders(1963)]{Flanders}  
Flanders, H. [1963], {\em  Differential Forms with Applications to the Physical Sciences},
\newblock Academic Press, 1963.

\bibitem[Gardner(1989)]{Gardner} 
Gardner, R. [1989], {\em The Method of Equivalence and its Applications},
\newblock SIAM, 1989.

\bibitem[Grossman(2000)]{Grossman}
Grossman, D.A. [2000], Torsion-free path geometries and integrable second order
ODE systems{\em Sel. math., New ser.} \textbf{6}, 399--442.

\bibitem[Guillemin and Sternberg(1980)]{Guillemin}
Guillemin, V., Sternberg, S. [1980], The moment map and collective motion, {\it Ann. of Phys.}
\textbf{1278}, 220--253.

\bibitem[Hamel(1949)]{Hamel} 
Hamel, G. [1949], {\em Theoretische Mechanik: Eine einheitliche Einführung in die gesamte Mechanik}, volume~57 of {\em Grundlehren der Mathematischen Wissenschaften}, revised edition,
\newblock Springer-Verlag, Berlin-New York, 1978. 

\bibitem[Haller and Rybicki(1999)]{Haller} 
Haller, S., Rybicki, T. [1999] On the group of diffeomorphisms preserving a locally conformal symplectic structure {\em Ann. Global Anal. Geom.} \textbf{17}, 475--502.

\bibitem[Haller and Rybicki(2001)]{Haller1}  
Haller, S., Rybicki, T. [2001] Symplectic reduction for locally conformal symplectic manifolds, {\em J. Geom. Phys.} \textbf{37} 262--271. 

\bibitem[Hertz(1899)]{Hertz}
Hertz, H. [1899], {\em
The principles of mechanics presented in a new form by Heinrich Hertz, with an introduction by H. von Helmholtz}, 
\newblock Macmillan,  London, New York,  1899. 

\bibitem[Hicks(1965)]{Hicks} 
Hicks, N.J. [1965], {\em Notes on Differential Geometry},  
\newblock Van Nostrand, 1965.

\bibitem[Hughen(1995)]{Hughen} 
Hughen, K. [1995], {\em The Geometry of SubRiemannian Three-Manifolds}, Ph.D. Thesis, 
\newblock Duke University, 1995.

\bibitem[Ibort, de L\'eon, Marrero and de Diego(1999)]{Ibort}
Ibort, A., de L\'eon, M., Marrero, J.C., de Diego, D. [1999], Dirac brackets in constraind
dynamics, {\em Fortschr. Phys.} \textbf{47}:5, 459--492.

\bibitem[Iliyev(1985)]{Iliyev}
Iliyev, Il., On the conditions for the existence of the reducing Chaplygin factor,
{\em P.M.M. USSR} \textbf{49}:2, 295--301.

\bibitem[Jovanovic(2003)]{Jovanovic} 
Jovanovic, B. [2003], Some multidimensional integrable cases of nonholonomic
rigid body dynamics, {\em Regular $\&$ Chaotic Dynamics} \textbf{8}:1, 125--132.

\bibitem[Kazarian, Montgomery and Shapiro(1997)]{Montgomery1} 
Kazarian, M. R., Montgomery, R., Shapiro, B. [1997],  Characteristic classes for the degenerations of two-plane fields in four dimensions, {\em Pacific J. Math. } \textbf{179}:2, 355--370.

\bibitem[Kobayashi and Nomizu(1963)]{Kobayashi} 
Kobayashi, S., Nomizu, K. [1963] {\em Foundations of Differential Geometry}, vol. 1,
\newblock J. Wiley, New York, London, 1963.

\bibitem[Kobayashi and Oliva(2003)]{KobayashiOliva}
Kobayashi, M.H., Oliva, W.M. [2003], A note on the conservation of energy and volume in the
setting of nonholonomic mechanical systems, {\em Qual. Theory Dyn. Systems}, to appear.

\bibitem[Koiller(1992)]{KoillerARMA}
Koiller, J. [1992], Reduction of some classical non-holonomic systems with symmetry, {\em Arch. Rational Mech. Anal.}  \textbf{118},  113--148.

\bibitem[Koiller,  Rodrigues and Pitanga(2001)]{Koiller1}
Koiller, J.,  Rodrigues, P.R., Pitanga, P. [2001],  Nonholonomic connections following \'Elie Cartan {\em Anais Academia Brasileira de Ciencias}  \textbf{73}:2, 165--190. 

\bibitem[Koiller and Rios(2001)]{KoillerRios} 
Koiller, J.,  Rios, P.M. [2001], Non-holonomic systems with symmetry allowing a conformally symplectic  reduction,    Proc. IV Int. Symp. of Hamiltonian Systems and Celestial Mechanics (Mexico 2001), edited by J. Delgado, E. A. Lacomba, E. P\'erez-Chavela,
\newblock to appear, Kluwer.

\bibitem[Koiller, Rios and Ehlers(2002)]{Koillerromp}  
Koiller, J., Rios, P.M., Ehlers, K. [2002], Moving frames for cotangent bundles, 
{\em Rep.   Math.  Phys.} \textbf{49}:2/3, 225--238.

\bibitem[Koiller(2003)]{Koillerreview}
Koiller, J. [2003], Analytical Mechanics, {\em Bull. (new series) of the AMS}, \textbf{40}:3, 
405--419.

\bibitem[Koon and Marsden(1997)]{Koon1}
Koon, W. S., Marsden, J.E. [1997], The Hamiltonian and Lagrangian approaches to the
dynamics of nonholonomic systems, {\em Rep. Math. Phys.} \textbf{40}, 21--62.

\bibitem[Koon and Marsden(1998)]{Koon}
Koon, W. S., Marsden, J.E. [1998], Poisson reduction for nonholonomic mechanical systems with symmetry, {\em Reports on Math. Phys.}, \textbf{42}, 101--134. 

\bibitem[Kozlov(2002)]{Kozlov}
Kozlov, V.V. [2002], On the integration theory of equations of nonholonomic mechanics,
{\em Reg. Chaot. Dyn.}, \textbf{7}:2, 161--176. 

\bibitem[Kummer(1981)]{Kummer}
Kummer, M. [1981], On the construction of the reduced phase space of a Hamiltonian
system with symmetry, {\em Indiana Univ. Math. J.}, \textbf{30}, 281--291.

\bibitem[Kupka and Oliva(2001)]{KupkaOliva}
Kupka, I., Oliva, W.M. [2001], The Non-Holonomic Mechanics, {\em J. Diff. Equations} \textbf{169}, 169--189.

\bibitem[Levi(1996)]{Levi}
Levi, M.[1996], Composition of rotations and parallel transport, {\em Nonlinearity} \textbf{9}, 413--419.
 
\bibitem[Lewis(1998)]{Lewis} 
Lewis, A. D. [1998], Affine connections and distributions with applications to nonholonomic mechanics,  {\em Rep. Math.  Phys.} \textbf{42}:(1/2), 135--164.

\bibitem[Marle(1995)]{Marle}
Marle, C.M. [1995], Reduction of constrained mechanical systems and stability of
relative equilibria, {\em Commun. Math. Phys.} \textbf{174}, 295–-318.


\bibitem[Marle(1998)]{Marle1}
Marle, C.M. [1998], Various approaches to conservative and nonconservative nonholonomic systems,  {\em Rep. Math. Phys.} \textbf{42}:1/2, 211--229.

\bibitem[Marle(2003)]{Marle2}
Marle, C.M. [2003], On symmetries and constants of motion in Hamiltonian systems with nonholonomic constraints, {\em Banach Center Publ.}, \textbf{59}, 223--242.

\bibitem[Marsden and Weinstein(1974)]{MWo}
Marsden, J.E., Weinstein, A. [2001],  Reduction of symplectic manifolds with symmetry,
{\em Rep. Math. Phys.} \textbf{5}, 121--130.

\bibitem[Marsden and Weinstein(2001)]{MW} 
Marsden, J.E., Weinstein, A. [2001],  Some comments on the history, theory, and applications
of symplectic reduction,  in Landsman, N.P., Pflaum, M., eds., {\em Quantization of
singular symplectic quotients}, 
\newblock Birkh\"auser, 2001.

\bibitem[Montgomery(1991)]{Montgomery0}
Montgomery, R.[1991], How Much Does the Rigid Body Rotate? A Berry's Phase from the 18th Century, {\em Am. J. Physics},
\textbf{59}:5, 394--398.

\bibitem[Montgomery(2002)]{Montgomery} 
Montgomery, R. [2002]   {\em A tour of subRiemannian geometries, their geodesics, and applications}, volume 91 of {\em  Mathematical surveys and monographs}, 
\newblock Providence, R.I.: American Mathematical Society, 2002. 

\bibitem[Moseley(2001)]{Moseley} 
Moseley, K. [2001] {\em The Geometry of SubRiemannian Engel Manifolds}, Ph.D. Thesis, 
\newblock Duke University, 2001.

\bibitem[Neimark and Fufaev(1972)]{NF} 
Neimark, J.I.,  N.~A.~Fufaev, N.A. [1972], {\em Dynamics of nonholonomic systems}
\newblock volume~33 of {\em AMS Translations of Mathematical Monographs}, Providence, 1972.

\bibitem[Oliva(2002)]{Oliva}
Oliva, W.M. [2002], {\em Geometric Mechanics}, volume 1798 of {\em Springer Lecture Notes
in Mathematics}
\newblock Springer Verlag, 2002.

\bibitem[Papastavridis(2002)]{Papastavridis} Papastavridis, J.G. [2002], {\em Analytical Mechanics:  a comprehensive treatise on the dynamics of constrained systems; for engineers, physicists and mathematicians},
\newblock   Oxford University Press, 2002. 
(Reviewed in \cite{Koillerreview}.)

\bibitem[Pars(1965)]{Pars}
Pars, L. A. [1965], {\em A treatise on analytical dynamics}
\newblock John Wiley,   New York, 1965. 

\bibitem[Poincar\'e(1901)]{Poincare} 
Poincar\'e, H. [1901], Sur une forme nouvelle des \'equations de la M\'ecanique,
{\em C. R. Acad. Sci. Paris} \textbf{132}, 369--371. 

\bibitem[Ramos(2004)]{Ramos}
Ramos, A. [2004], Poisson structures for reduced non-holonomic systems, preprint,
arXiv: math-ph/0401054 v1 .
 
\bibitem[Salamon(1989)]{Salamon} 
Salamon, S. [1989], {\em Riemannian geometry and holonomy groups},
volume~201 of {\em Pitman Research Notes in Mathematics},
\newblock Longman Scientific and Technical, Essex, 1989.

\bibitem[Schneider(2002)]{Schneider}
Schneider, D., Nonholonomic Euler-Poincare Equations and Stability in Chaplygin's Sphere, {\em Dyn. Syst.: an Intern. J.}, \textbf{17}:2, 87--130. 

\bibitem[Sharpe(1997)]{Sharpe} 
Sharpe, R.W. [1997] {\em Differential Geometry: Cartan's
generalization of Klein's Erlangen program}, volume~166 of {\em Graduate Texts in Math.}
\newblock Springer-Verlag, 1997. 

\bibitem[\'Sniatycki(1998)]{Sniatycki98}
\'Sniatycli, J.[1998], Nonholonomic Noether theorem and reduction of symmetries,
{\em Rep. Math. Phys.} \textbf{42}:1/2, 5--23.

\bibitem[S\'niatycki(2001)]{Sniatyckisingular}
S\'niatycki, J.[2001], Almost Poisson spaces and nonholonomic singular reduction,
{\em Rep. Math. Phys.} \textbf{48}:1/2, 235--248.

\bibitem[\'Sniatycki(2002)]{Sniatycki2002}
\'Sniatycki, J.[2002], The momentum equation and the second order differential condition,
{\em Rep. Math. Phys.} \textbf{213}, 371--394.

\bibitem[Sommerfeld(1952)]{Sommerfeld} 
Sommerfeld, A. [1952], {\em  Lectures on Theoretical Physics: Mechanics}
\newblock   Academic Press, 1952. 

\bibitem[Stanchenko(1985)]{Stanchenko}
Stanchenko, S.V. [1985], Non-holonomic Chaplygin systems, {\em P.M.M. USSR} \textbf{53}:1,
11--17.

\bibitem[Tavares(2002)]{Tavares}
Tavares, J.N. [2002], About Cartan geometrization of non holonomic mechanics, preprint,
Centro de Matem\'atica da Universidade do Porto,   www.fc.up.pt/cmup .

\bibitem[Mashke and van der Schaft(1994)]{MS} 
van der Schaft, A., Mashke, B.M. [1994], On the hamiltonian formulation of nonholonomic mechanical systems, {\em Rep. Math. Phys.} \textbf{34}, 225 -- 233.

\bibitem[Vaisman(1985)]{Vaisman} 
Vaisman, I., Locally conformal symplectic manifolds. 
{\em Internat. J. Math. Math. Sci.} \textbf{8}:3, 521--536. 

\bibitem[Vershik and Gerhskovich(1994)]{Vershik} 
Vershik, A.M., Gerhskovich, V. [1994] {\em Nonholonomic Dynamical Systems, Geometry of Distributions and Variational Problems}, in {\em Dynamical Systems VII} ed. V.I. Arnol'd and S.P. Novikov, vol. 16 of {\em the Encyclopedia of Mathematical Sciences series}, 
\newblock Springer-Verlag, NY, 1994.

\bibitem[Vershik and Fadeev(1981)]{VershikFadeev} 
Vershik, A.M., Fadeev, L.D. [1981],
Lagrangian mechanics in invariant form, {\em Selecta Math. Sov.} \textbf{1}:4, 339--350.
           
\bibitem[Veselov and Veselova(1986)]{Veselovs1}
Veselov, A. P., Veselova, L. E.,  Flows on Lie groups with a nonholonomic constraint and integrable non-Hamiltonian systems. (Russian), {\em Funktsional. Anal. i Prilozhen.} \textbf{20}:4, 65--66. 
English translation: {\em Functional Anal. Appl.} \textbf{20}:4,308--309.

\bibitem[Veselov and Veselova(1988)]{Veselovs}
Veselov, A. P., Veselova, L. E., Integrable nonholonomic systems on Lie groups. (Russian){\em Mat. Zametki} \textbf{44}:5   604--619, 701; translation in {\em Math. Notes} \textbf{44}:5/6, 810--819 (1989). 

\bibitem[Wade(2000)]{AissaWade} 
Wade, A. [2000], Conformal Dirac structures, {\em Lett. Math. Phys.}
\textbf{53},  331--348.

\bibitem[Warner(1971)]{Warner}  
Warner, F. [1971] {\em Foundations of Differentiable Manifolds and Lie Groups}, 
\newblock Scott Foresman, 1971.

\bibitem[Weber(1986)]{Weber}
Weber, R.W.[1986], Hamiltonian systems with constraints and their meaning in mechanics,
{\it Arch. Ratl. Mech. Anal.} \textbf{91}, 309--335.

\bibitem[Whittaker(1937)]{Whittaker}
Whittaker, E. T. [1988], {\em A treatise on the analytical dynamics of particles and rigid bodies. With an introduction to the problem of three bodies. Reprint of the 1937 edition}
\newblock Cambridge University Press, Cambridge, 1988.

\bibitem[Zenkov(1995)]{Zenkov}
Zenkov, D. V.[1995], The geometry of the Routh problem,
{\em J. Nonlinear Sci.} \textbf{5}:6, 503--519.

\bibitem[Zenkov and Bloch(2000)]{ZenkovBloch}
Zenkov, D.V., Bloch, A.M.[2000], Dynamics of the $n$-dimensional Suslov problem, 
{\em J. Geom. Phys.} \textbf{34}:2,  121--136.

\bibitem[Zenkov and Bloch (2003)]{Blochnonlinearity}
Zenkov, D.V., Bloch, A.M. [2003],   Invariant measures of nonholonomic flows with
internal degrees of freedom, {\em Nonlinearity} \textbf{16}, 1793--1807.

\end{thebibliography}
}
\end{document}